\documentclass[11pt]{article}
\usepackage{html,amsmath,amssymb,amsfonts,epsfig,cancel,cite,longtable,caption,siunitx,array,color,booktabs,setspace}

\newlength{\xtrawidth}
\newlength{\xtraheight}
\newcommand{\be}{\begin{equation}}
\newcommand{\ee}{\end{equation}}
\newcommand{\beq}{\begin{equation}}
\newcommand{\eeq}{\end{equation}}
\newcommand{\ba}{\begin{array}}
\newcommand{\ea}{\end{array}}
\newcommand{\bea}{\begin{eqnarray}}
\newcommand{\eea}{\end{eqnarray}}
\newcommand{\bean}{\begin{eqnarray*}}
\newcommand{\eean}{\end{eqnarray*}}

\newcommand{\IC}{\mathbb{C}}
\newcommand{\IP}{\mathbb{P}}
\newcommand{\IR}{\mathbb{R}}
\newcommand{\IH}{\mathbb{H}}

\newcommand{\IZ}{\mathbb{Z}}
\newcommand{\cO}{{\cal O}}
\newcommand{\cN}{{\cal N}}

\newcommand{\cA}{{\cal A}}

\newcommand{\calO}{\mathcal{O}}

\def\fnote#1#2{\begingroup\def\thefootnote{#1}\footnote{#2}
     \addtocounter{footnote}{-1}\endgroup}

\newcommand{\cicy}[2]{\begin{matrix} #1\end{matrix}\!\left[\begin{matrix}#2 \end{matrix}\right]}

\newcommand*{\myalign}[2]{\multicolumn{1}{#1}{#2}}
\newcommand{\varstr}[2]{\vrule height #1 depth #2 width0pt}

\numberwithin{equation}{section}

\begin{document}

\vspace{1cm}

\title{{\LARGE \bf The Moduli Space of Heterotic Line Bundle Models:\\ a Case Study for the Tetra-Quadric}}

\vspace{2cm}

\author{
Evgeny I. Buchbinder${}^{1}$,
Andrei Constantin${}^{2}$,
Andre Lukas${}^{2}$
}

\date{}
\maketitle
{\setstretch{1.3}
\begin{center} {\small ${}^1${\it The University of Western Australia, \\
35 Stirling Highway, Crawley WA 6009, Australia\\[0.3cm]
       ${}^2$Rudolf Peierls Centre for Theoretical Physics, Oxford
       University,\\
       $~~~~~$ 1 Keble Road, Oxford, OX1 3NP, U.K.}}\\

\fnote{}{evgeny.buchbinder@uwa.edu.au, a.constantin1@physics.ox.ac.uk, lukas@physics.ox.ac.uk} 

\end{center}
}

\vspace{12pt}
\begin{abstract}
\noindent It has recently been realised that polystable, holomorphic sums of line bundles over smooth Calabi-Yau three-folds provide a fertile ground for heterotic model building. Large numbers of phenomenologically promising such models have been constructed for various classes of Calabi-Yau manifolds. In this paper we focus on a case study for the tetra-quadric - a Calabi-Yau hypersurface embedded in a product of four $\IC\IP^1$ spaces. We address the question of finiteness of the class of consistent and physically viable line bundle models constructed on this manifold. Further, for a specific semi-realistic example, we explore the embedding of the line bundle sum into the larger moduli space of non-Abelian bundles, both by means of constructing specific polystable non-Abelian bundles and by turning on VEVs in the associated low-energy theory. In this context, we explore the fate of the Higgs doublets as we move in bundle moduli space. 
The non-Abelian compactifications thus constructed lead to $SU(5)$ GUT models with an additional global $B-L$ symmetry. The non-Abelian compactifications inherit many of the appealing phenomenological features of the Abelian model, such as the absence of dimension four and dimension five operators triggering fast proton decay. 
\end{abstract}

\thispagestyle{empty}
\setcounter{page}{0}

\newpage

{\setstretch{1.2}
\thispagestyle{empty}
\tableofcontents
}

%

\section{Introduction}
Smooth Calabi-Yau compactifications of the heterotic string and M-theory represent one of the classic and most-developed avenues from string theory to low energy physics \cite{Candelas:1985en}. Despite the great interest received over the years, however, this approach, until recently, has led to only a relatively  small number of models that satisfy the most basic phenomenologically requirements, such as exhibiting the Standard Model particle content without any exotics \cite{Braun:2005ux, Braun:2005bw, Bouchard:2005ag, Anderson:2009mh, Braun:2011ni}. The scarceness of such models, owing to the considerable mathematical complexity involved in the analysis of the compactification geometry, has made it difficult to address more detailed phenomenological questions, such as proton decay or fermion masses, in a meaningful way.

Recently, this situation has changed.  In a series of publications \cite{Anderson:2011ns, Anderson:2012yf, Anderson:2013xka, He:2013ofa}, a promising class of $E_8\times E_8$ heterotic compactifications, based on line bundle sums, has been proposed and analyzed. These heterotic line bundle models are based on rank four or five Abelian vector bundles over smooth Calabi-Yau three-folds with non-trivial fundamental group. Using line bundle sums, rather than bundles with non-Abelian structure groups, comes with a number of far-reaching advantages. For one, split bundles lead to the presence of additional, normally anomalous $U(1)$ symmetries which constrain the structure of the low-energy theory and can have interesting phenomenological implications. Further, the split nature of the bundle facilitates an efficient algorithmic search for physical models and large numbers of promising examples can be found in this way.
Specifically, in Refs.~\cite{Anderson:2011ns, Anderson:2012yf} over 200 $SU(5)$ GUT models were constructed on discrete quotients of complete intersection Calabi-Yau manifolds. All these have precisely three generations of GUT families, no anti-families, at least one $\mathbf{5} - \overline{\mathbf{5}}$ pair of Higgs fields and no other matter charged under the GUT group. After forming quotients by the freely-acting discrete symmetries and including Wilson lines, these models led to about 2000 models with the MSSM matter field spectrum plus one or more pair of Higgs doublets. In Ref.~\cite{Anderson:2013xka}, this preliminary search for standard-like models was extended to an exhaustive scan which led to some $35,000$ $SU(5)$  GUT models over the same class of Calabi-Yau manifolds. Finally, in \cite{He:2013ofa}, over $100$ $SU(5)$ models and about $29,000$ $SO(10)$ GUT models were constructed over $14$~Calabi-Yau hypersurfaces embedded in toric varieties. These large numbers are indicative of the huge potential of the line bundle construction. 

The present paper will be studying heterotic line bundle models using a complementary approach. Rather than scanning over classes of Calabi-Yau manifolds and large numbers of bundles, we will be focusing on a specific Calabi-Yau manifold, the tetra-quadric hypersurface in $\left(\IC\mathbb{P}^1\right)^{\!\times 4}$, for an in-depth analysis of some aspects of heterotic line bundle models. In particular, we will use the present case study to elaborate on three, related points. Firstly, it was noticed in \cite{Anderson:2013xka} that the number of viable models reaches a certain saturation limit after repeatedly increasing the range of integers defining the line bundles. 
While this observation was made computationally, it is less clear how to prove finiteness of the class of relevant vector bundles -- essentially poly-stable line bundle sums with fixed total Chern class. We revisit this question here and present various partial results including a finiteness proof which relies on excluding the regions in K\"ahler moduli space close to the boundary; in physical terms this essentially corresponds to the supergravity approximation. 

The second objective of the paper is to present in detail a line bundle model on the tetra-quadric manifold that has modestly attractive phenomenological properties. The model exhibits a superpotential that leads to a rank two up quark mass matrix and, while a rank one matrix may be preferably at this level, this means a perturbative and generically large top Yukawa coupling is present. The down quark and lepton Yukawa matrices are entirely zero at the perturbative level so, for a realistic model, they would have to be generated non-perturbatively. Further, all dimension four and five operators which can lead to fast proton decay are forbidden. This model serves as a starting point for the subsequent analysis of the bundle moduli space.

Finally, perhaps the most important issue is to understand how a given line bundle model is embedded into the larger moduli space of non-Abelian bundles and we will study this question for the aforementioned model on the tetra-quadric. Methodically, this problem can be approached in two complementary ways.  From a  fundamental point of view, one can attempt to construct non-Abelian bundles which split to the given line bundle sum at a particular locus in bundle moduli space. On the other hand, in the context of the four-dimensional effective theory, the analogous process can be carried out by giving VEVs to the bundle moduli. For our example model, we will consider both approaches as well as the relation between them. Similar studies have been undertaken in \cite{Anderson:2010ty, Kuriyama:2008pv, Anderson:2010tc}. An important question in this context concerns the fate of the Higgs doublet pair. As a vector-like pair, the Higgs is not automatically protected from acquiring a mass as one moves in moduli space. In fact, for most of the semi-realistic models constructed to date, the Higgs doublets are massless only on a specific sub-locus in moduli space and receive a typically super-heavy mass elsewhere. For the line bundle model, we have -- by construction -- a massless Higgs pair at the Abelian locus, but this is not guaranteed to remain massless as we move into the non-Abelian part of the moduli space. Clearly, the conditions under which Higgs doublets are part of the low-energy spectrum, in low-energy parlance referred to as the $\mu$-problem,  is of vital importance for string model building and we will study this question for our example model.

These seemingly dissimilar questions run along a common thread and illustrate several problems which arise when studying the moduli space of heterotic line bundle models.  
Coming from afar, we identify -- for our tetra-quadric example manifold -- a substantial, though finite, number of points in this moduli space representing Abelian bundles which lead to phenomenologically promising 
low-energy theories. In the second step, Section~\ref{sec:model}, we focus on one of these points and, in the last part, Section~\ref{sec:monads}, we ``zoom out" again, 
this time in order to explore the space of non-Abelian deformations around the chosen point.

The non-Abelian bundles discussed in Section~\ref{sec:monads} have structure group $SU(4) \times U_X(1)$ leading to $SU(5)$ GUT models with an additional global $U_X(1)$ symmetry which can be seen as a remnant of the four $U(1)$ symmetries present at the Abelian locus. Combined with hypercharge, this $U_X(1)$ leads to a $B-L$ symmetry. Furthermore, these models have a number of appealing phenomenological features such as a certain hierarchy of Yukawa couplings and the absence of dimension four and dimension five operators leading to a fast proton decay. In part, these operators are forbidden by the surviving $U_X(1)$ symmetry. However, it is well known that operators of the form ${\bf 10}\,{\bf 10}\,{\bf 10}\,\overline{\bf 5}$ are invariant under the $U_X(1)$ symmetry. Yet, in our $SU(5)\times U_X(1)$ models these operators are absent because of the presence of additional $U(1)$ symmetries which appear at the Abelian locus. Something similar happens with the Higgs multiplets. By construction we have a massless pair of Higgs doublets at the Abelian locus but, as we will see, the $U(1)$ symmetries also forbid a $\mu$-term with singlet insertions. This means that the Higgs doublets remain massless as we continue away from the Abelian locus to an $SU(5)\times U_X(1)$ model. These examples demonstrate the power and the advantage of the present approach: we start with an Abelian model which is easier to construct and analyse, but the symmetries which arise at the Abelian locus lead to some degree of control as to which couplings will or will not appear as we move into the non-Abelian part of the moduli space. 

The discussion runs on two, even three, levels. On one hand, we have the high energy theory, described in terms of the compactification data: a Calabi-Yau three-fold supplied with a holomorphic, poly-stable bundle. On the other hand, this geometrical set-up leads to a four-dimensional, low-energy supersymmetric GUT, whose gauge group can further be broken to that of the Standard Model. Frequently, we switch back and forth between the high-energy and 
the effective GUT. The breaking to the Standard Model is only discussed briefly in Section~\ref{sec:model}, simply to illustrate the virtues of the chosen line bundle model and will be presented in detail elsewhere.

\section{Heterotic line bundle models}\label{sec:lbs}
We begin by reviewing the construction of smooth heterotic compactifications with Abelian bundles. The line bundle construction has been extensively discussed in Refs.~\cite{Anderson:2011ns, Anderson:2012yf} and, below, we summarise its most important features, including the derivation of the GUT spectrum. Knowledge of the GUT spectrum will be crucial for the comparison between Abelian and non-Abelian compactifications in Section~\ref{sec:monads}.

\subsection{Construction}\label{sec:construction}
In the rest of the paper we will discuss $E_8\times E_8$ heterotic line bundle models leading to $SU(5)$ GUT models. For this purpose, we choose a Calabi-Yau three-fold, $X$, with a freely-acting discrete symmetry $\Gamma$ and a vector bundle $V\rightarrow X$ which is given by the sum of line bundles
\begin{equation}
V = \bigoplus_{a=1}^5 L_a\; .
\end{equation}
In order for $V$ to have structure group $S\left(U(1)^5\right)$, such that we are able to use the embedding $S\left(U(1)^5\right)\subset SU(5)\subset E_8$, we demand that $c_1(V)=0$. In practice, we can choose an integral basis, $\{J_i\}$ of the second cohomology of $X$ and characterise line bundles by their first Chern class, that is, write $L_a={\cal O}({\bf k}_a)$ if $c_1(L_a)=k_a^iJ_i$. Here the indices $i,j,\ldots $ run from $1$ to $h^{1,1}(X)$. The above line bundle sum is then specified by an $h^{1,1}(X)\times 5$ matrix, $(k_a^i)$, of integers and the vanishing of the first Chern class, $c_1(V)=0$, translates into the condition
\vspace{-4pt}
\begin{equation}
 \sum_{a=1}^5{\bf k}_a=0\; . \label{c10}
 \vspace{-4pt}
\end{equation} 
The second Chern class and the index of such line bundle sums are given by
 \vspace{-4pt}
\begin{equation}
 c_2(V)=-\frac{1}{2}d_{ijk}\sum_{a=1}^5k_a^ik_a^j\;,\quad\quad
 {\rm ind}(V)=\frac{1}{6}d_{ijk}\sum_{a=1}^5k_a^ik_a^jk_a^k\; ,\label{c2ind}
  \vspace{-4pt}
\end{equation}
where the triple intersection numbers $d_{ijk}$ of $X$ are defined by
\begin{equation}
 d_{ijk}=\int_XJ_i\wedge J_j\wedge J_k\; ,
\end{equation}
 as usual. For a consistent heterotic compactification we have to ensure that the anomaly condition 
 \vspace{-4pt}
 \begin{equation}
  {\rm ch}_2(V) +{\rm ch}_2(V')-{\rm ch}(TX)=[W]\; ,\label{anomgen}
   \vspace{-4pt}
 \end{equation}
 can be satisfied, where $V'$ is the bundle in the other (hidden) $E_8$ sector and $[W]$ is the (Poincar\'e dual of the) class of a holomorphic curve $W$ wrapped by a five-brane. A practical way to ensure this condition can be satisfied without having to explicitly construct the hidden sector is to demand that~\footnote{We are assuming here that $c_1(V)=0$, so that ${\rm ch}_2(V)=-c_2(V)$.} 
 \vspace{-4pt}
\begin{equation}
 c_2(TX)-c_2(V)\in \mbox{Mori cone of }X\; . \label{anom}
\end{equation}  
Then the anomaly condition~\eqref{anomgen} can always be satisfied by a suitable choice of five-brane (although other configurations which involve a non-trivial hidden bundle are normally possible as well). 

Finally, we need to guarantee that the observable bundle is poly-stable with slope zero, so that supersymmetry is preserved by the gauge fields. Line bundles are automatically stable, so what remains to be checked is that all line bundles have vanishing slope
 \vspace{-4pt}
\begin{equation}
 \mu(L_a)\,\equiv\int_Xc_1(L_a)\wedge J\wedge J=d_{ijk}\,k_a^i\,t^j\,t^k\stackrel{!}{=}0\label{slope0}
  \vspace{-8pt}
\end{equation}
for a common locus in K\"ahler moduli space, parametrized by $J=t^iJ_i$ with moduli $t^i$. The relative simplicity of this condition, as opposed to the condition of stability for non-Abelian bundles, is one of the major technical advantages of line bundle models. 

We note that a line bundle $L$ (other than the trivial bundle) with vanishing slope, $\mu(L)=0$, has vanishing zeroth and third cohomology, $H^0(X,L)=H^3(X,L)=0$ so that
\begin{equation}
 {\rm ind}(L)=-h^1(X,L)+h^1(X,L^{^*})\; .
\end{equation}  
\subsection{The Spectrum}
For a bundle structure group $S(U(1)^5)\subset SU(5)\subset E_8$ the low-energy gauge group, given by the commutant of the structure group within $E_8$, is $SU(5)\times S\left(U(1)^5\right)$. The matter multiplets present in the four-dimensional theory can be obtained by decomposing the adjoint ${\bf 248}_{E_8}$ of $E_8$ under the $SU(5)\times S\left(U(1)^5\right)$ sub-group which leads to 
\begin{equation}
 {\bf 10}_a\; ,\quad \overline{\bf 10}_a\; ,\quad \overline{\bf 5}_{a,b}\;,\quad {\bf 5}_{a,b}\; ,\quad {\bf 1}_{a,b}\; ,
\end{equation} 
Here the number indicates the $SU(5)$ representation and the indices $a,b,\ldots =1,\dots 5$ indicate which of the five $U(1)$ symmetries the multiplet is charged under. Specifically, the ${\bf 10}_a$ ($\overline{\bf 10}_a$) multiplets carry charge $+1$ ($-1$) under the $a^{\rm th}$ $U(1)$ symmetry while being uncharged under the others. The $\overline{\bf 5}_{a,b}$ (${\bf 5}_{a,b}$) multiplets carry charge $+1$ ($-1$) under $U(1)$ charges $a$ and $b$ while the singlets ${\bf 1}_{a,b}$ carry charge $+1$ under the $a^{\rm th}$ and charge $-1$ and the $b^{\rm th}$ $U(1)$. The multiplicity of each of these multiplets can be computed from line bundle cohomology, as summarised in Table~\ref{spectrum}.
\begin{table}[!h]
\vspace{12pt}
\begin{center}
\begin{tabular}{|l|l|l|l|l|l|l|l|}
\hline
\varstr{14pt}{9pt} repr. & cohomology & total number & required for MSSM \\ \hline\hline
\varstr{14pt}{9pt} $~{\bf 1}_{a,b}$ & $H^1(X, L_a \otimes L_b^{^*})$  &  $\sum_{a,b} h^1(X, L_a \otimes L_b^{^*}) = h^1(X, V \otimes V^{^*})$ & \;\;\;\;\; - \\ \hline
\varstr{14pt}{9pt} $~{\bf 5}_{a,b}$ & $H^1(X, L_a^{^*} \otimes L_b^{^*})$  & $\sum_{a<b} h^1(X, L_a^{^*} \otimes L_b^{^*}) =h^1(X, \wedge^2 V^{^*}) $ & \;\;\;\;\;$n_h$\\ \hline
\varstr{14pt}{9pt} $~{\bf \overline{5}}_{a,b}$ & $H^1(X, L_a \otimes L_b)$  & $\sum_{a<b} h^1(X, L_a \otimes L_b) =h^1(X, \wedge^2 V) $ & \;\;\;\;\;$3 |\Gamma| + n_h$\\ \hline
\varstr{14pt}{9pt} $~{\bf 10}_{a}$ &$H^1(X, L_a)$ & $\sum_a h^1 (X,L_a) = h^1 (X,V)$& \;\;\;\;\;$3 | \Gamma|$\\ \hline
\varstr{14pt}{9pt} $~{\bf  \overline{10}}_{a}$ & $H^1(X, L_a^{^*})$ & $\sum_a h^1(X,L_a^{^*}) = h^1(X,V^{^*})$&\;\;\;\;\; 0
\\ \hline 
 \end{tabular}
 \vskip 0.4cm
\parbox{16.7cm}{\caption{\it\small The spectrum of $SU(5)$ GUT models derived from the heterotic line bundle construction. In the final column, $|\Gamma|$~stands for the order of the fundamental group of $X$ and  $n_h$ represents the number of $\mathbf{5}-\overline{\mathbf{5}}$ Higgs fields.}\label{spectrum}}
 \end{center}
 \vspace{-14pt}
 \end{table}
The cohomology of line bundles is usually easier to compute than that of non-Abelian bundles and this constitutes another major technical advantage of line bundle models. The phenomenological requirements on the GUT particle spectrum -- essentially the three-family constraint plus having an additional ${\bf 5} - \overline{\bf 5}$ pair to account for the Higgs doublets -- are summarized in the last column of Table~\ref{spectrum}.

In order to arrive at a standard-like model, we need a freely-acting symmetry $\Gamma$ on $X$, with order~$|\Gamma|$, which can be lifted to the bundle $V$, that is, the bundle $V$ needs to have a $\Gamma$-- equivariant structure. Then, performing the quotient by $\Gamma$ and including a Wilson line in the hypercharge direction will break the GUT group to the Standard Model group times $S(U(1)^5)$. These additional $U(1)$ symmetries are usually Green-Schwarz anomalous with super-heavy associated gauge bosons and, therefore, do not constitute a phenomenological problem. Upon quotienting by $\Gamma$, the number $3|\Gamma|$ of ${\bf 10}\oplus\overline{\bf 5}$ families which we have required for our GUT models, automatically become $3$ standard model families. From the additional $\overline{\bf 5}-{\bf 5}$ multiplets we should keep one pair of Higgs doublets and ensure that all Higgs triplets are projected out. This can frequently be achieved by a suitable choice of equivariant structure and Wilson line. Then, we have a standard model charged spectrum precisely as in the MSSM plus additional moduli fields -- the bundle moduli ${\bf 1}_{a,b}$ and gravitational moduli -- which are uncharged under the standard model group. Our experience is that many such models can be found relatively easily and in this paper we focus on the examples on the tetra-quadric.

{\setstretch{1.14}
\section{Heterotic line bundle models on the tetra-quadric}
In this section, we focus on the tetra-quadric manifold, discuss its specific properties and present the scan for phenomenologically interesting models on this manifold. 

\subsection{The tetra-quadric}
A detailed discussion of the tetra-quadric, particularly of its K\"ahler cone, is provided in Appendix~\ref{app:KahlerCone}. Here we summarise the most important points. Tetra-quadric Calabi-Yau hypersurfaces are embedded in a product of four $\IC\IP^1$ spaces, defined as the zero locus of some homogeneous polynomial that is quadratic in the homogeneous coordinates of each $\IC\IP^1$ space. Manifolds in this class have Euler number $\eta = - 128$ and Hodge numbers $h^{1,1}(X)=4$ and $h^{2,1}(X)=68$. This information is summarised by the following configuration matrix:
\begin{equation}
X~=~~
\cicy{\IC\IP^1 \\   \IC\IP^1\\ \IC\IP^1\\ \IC\IP^1}
{ ~2 \!\!\!\!\\
  ~2\!\!\!\! & \\
  ~2\!\!\!\! & \\
  ~2\!\!\!\!}_{-128}^{4,68}\
\end{equation}
At certain loci in the complex structure moduli space, the tetraquadric hypersurface admits free actions of finite groups of orders $|\Gamma|=2, 4, 8,16$. Specifically, these groups are $\Gamma= \IZ_2,\,\IZ_2\times \IZ_2,\, \IZ_4$, $\IZ_2\times \IZ_4,\,\IZ_8,\,\IH,\,\IZ_4\times \IZ_4,\,\IZ_4 \rtimes \IZ_4,\,\IZ_8\times \IZ_2,\,\IZ_8\rtimes \IZ_2,\,\IH\times \IZ_2$. Being at one or another of these special loci corresponds to different choices of coefficients for the monomials composing the defining polynomial, as discussed in Refs.~\cite{Candelas:2008wb, Candelas:2010ve}. In other words, saying that the tetraquadric manifold $X$ admits free quotients by a finite group $\Gamma$ implies a partial fixing of the complex structure of $X$. In due course, when we consider line bundle models on the tetra-quadric, some of the K\"ahler moduli will also be fixed by virtue of the slope zero conditions~\eqref{slope0}. 

The tetra-quadric is ``favourable" in the sense that its entire second cohomology is spanned by the K\"ahler forms $J_1,\ldots ,J_4$ of the four $\IC\IP^1$ factors, restricted to the hypersurface. Its cone of K\"ahler forms $J=\sum_{i=1}^4t^iJ_i$ is given by
\beq
C_{{\bf t}} =\left\{ {\bf t} \in \IR^4\ \left|\ t^i\geq 0,\, 1\leq i\leq 4\right. \right\} \label{kc}
\eeq
and the triple intersection numbers have the following simple form
\beq
d_{ijk} = \int_X J_i\wedge J_j\wedge J_k = \begin{cases} 2 & \mbox{ if } i\neq j, j\neq k \\ 0 &\mbox{ otherwise } \end{cases}\; .\label{tqisec}
\eeq
This leads to the volume form
\begin{equation}
\kappa  = 12\left( t_1t_2t_3+t_1t_2t_4+t_1t_3t_4+t_2t_3t_4\right)\; .
\end{equation}
The second Chern class of the tangent bundle of the tetra-quadric, in the basis $\{\nu^i\}$ of four-forms dual to $J_i$, is given by
\begin{equation}
 c_2(TX)=(24,24,24,24)\; . \label{tqc2}
\end{equation} 
The Mori cone corresponds to all positive linear combinations of $\nu^i$.

\subsection{Line bundle models}
To construct line bundle models on the tetra-quadric we follow the general discussion in Section~\ref{sec:lbs}. A~line bundle sum is specified by a $4\times 5$ integer matrix $(k_a^i)$ subject to the constraint~\eqref{c10} to ensure the vanishing of the first Chern class of the bundle. In Appendix~\ref{app:lbtopology} we provide a more comprehensive account of topological identities of interest for line bundles on the tetra-quadric manifold. Here we focus on the most important quantities, starting with the second Chern class which, from Eqs.~\eqref{c2ind} and \eqref{tqisec}, is given by
\begin{equation}
 c_{2i}(V)=-2\sum_{a=1}^5(k^2_ak^3_a+k^2_ak^4_a+k^3_ak^4_a,k^1_ak^3_a+k^1_ak^4_a+k^3_ak^4_a,k^1_ak^2_a+k^1_ak^4_a+k^2_ak^4_a,k^1_ak^2_a+k^1_ak^3_a+k^2_ak^3_a) \label{c2}
\end{equation}
relative to the basis $\{\nu^i\}$. Then,  from Eq.~\eqref{tqc2}, the anomaly cancellation condition~\eqref{anom} becomes
\begin{equation}
 c_{2i}(V)\leq 24\; . \label{tqanom}
\end{equation}
Again, from Eqs.~\eqref{c2ind} and \eqref{tqisec}, the index can be computed as
\begin{equation}
 {\rm ind}(V)=2\sum_{a=1}^5\left(k^1_ak^2_ak^3_a+k^1_ak^2_ak^4_a+k^1_ak^3_ak^4_a+k^2_ak^3_ak^4_a\right)\; .
\end{equation}
Defining
\begin{equation}
 (\kappa_i)=4(t_2t_3+t_2t_4+t_3t_4,t_1t_3+t_1t_4+t_3t_4,t_1t_1+t_1t_4+t_2t_4,t_1t_2+t_1t_3+t_2t_3)\label{kappai} 
\end{equation} 
the slope zero conditions~\eqref{slope0} translate into
\begin{equation}
\mu(L_a)=\kappa_ik^i_a\stackrel{!}{=}0\; . \label{tqslope0}
\end{equation}
These conditions need to be satisfied simultaneously  for all $a=1,\ldots ,5$ somewhere in the interior of the K\"ahler moduli space, so for moduli values $t^i>0$. Of course, the vanishing of the first Chern class~\eqref{c10} ensures that at most four of these conditions are independent. Indeed, for a non-trivial solution of the Eqs.~\eqref{tqslope0} at most three of these condition can be independent, so a necessary condition for having a solution with vanishing slope is that the matrix $(k_a^i)$ of line bundle integers satsifies
\begin{equation}
 {\rm rank}(k_a^i)\leq 3\; . \label{rkcons}
\end{equation} 
Of the four additional $U(1)$ symmetries, $4-{\rm rank}(k^i_a)$ are non-anomalous and, hence, have massless gauge bosons at the line bundle locus. From~\eqref{rkcons} this means there is at least one such non-anomalous $U(1)$ symmetry present for models on the tetra-quadric. Such a $U(1)$ symmetry is phenomenologically unwanted and it can be spontaneously broken by giving VEVs to the singlets ${\bf 1}_{a,b}$. This corresponds to moving into the non-Abelian part of the moduli space, something we will explore in detail later on in the paper. 

\begin{table}[!h]
\vspace{12pt}
\begin{center}
\begin{tabular}{| c || c | c | c | c | c |}
\hline
\myalign{| c||}{\varstr{21pt}{16pt}$\ \ \ \  |\Gamma| $ $\ \ \ \ $} &
\myalign{m{3.2cm}|}{$\ $ GUT models} &
\myalign{m{3.5cm}|}{ $\ \ \ \ $ no $ \overline{\mathbf{10}}$ multiplets$\ \ \ $ }&
\myalign{m{3.5cm}|}{$\ \ \ \ \ \ \ $ no $ \overline{\mathbf{10}}\,$s  and $\ \ \ \ \ \ \ $ at least one $\mathbf{5}-\overline{\mathbf{5}}$ pair} & 
$\ \ \ k_{\text{max}}\ \ \ $
\\ \hline\hline
\varstr{14pt}{8pt} 2 & 10 & 10 & 8 & 4 \\
 \hline
\varstr{14pt}{8pt} 4 & 58 & 53 & 46 & 5 \\
 \hline
\varstr{14pt}{8pt} 8 & 64 & 52 & 36 & 7 \\
 \hline
\varstr{14pt}{8pt} 16 & 5 & 5 & 4 & 6 \\
 \hline
 \end{tabular}
 \vskip 0.4cm
\parbox{16.7cm}{\caption{\it\small Statistics on the number of models on the tetraquadric manifold. The second column gives the number of viable GUT models consistent with the phenomenologically required values of the indices, for each group order $|\Gamma|$. The third and fourth columns give the number of such models which, in addition, satisfy the cohomology constraints specified. All models satisfy $|k_a^i|\leq k_{\rm max}$ with the smallest such value $k_{\rm max}$ given in the last column.}\label{tqmodels}}
 \end{center}
 \vspace{-12pt}
 \end{table}

The most basic physical constraints on the spectrum are the ones which can be formulated in terms of the index. Most importantly, we have the three-family constraint ${\rm ind}(V)\stackrel{!}{=}-3|\Gamma|$ for the ${\bf 10}$ multiplets which, for $SU(5)$-bundles, implies that the three-family constraint ${\rm ind}(\wedge^2 V)=-3|\Gamma|$ for the $\overline{\bf 5}$ multiplets is automatically satisfied. In addition, we require that $-3|\Gamma|\leq {\rm ind}(L_a\otimes L_b)\leq 0$ for all $a<b$, where the lower limit is just as to not exceed the three-family bound and the upper limit is to avoid chiral ${\bf 5}_{a,b}$ multiplets.
At the more sophisticated level of cohomology, we should demand the absence of $\overline{\bf 10}$ multiplets, that is, $h^1(X,V^{^*})\stackrel{!}{=}0$ and the presence of at least one ${\bf 5}-\overline{\bf 5}$ pair, that is, $h^1(X,\wedge^2V^{^*})\stackrel{!}{>}0$. These constraints can be checked using the results from Appendix~\ref{app:tqcoh} where we present an explicit formula for computing line bundle cohomology on the tetra-quadric manifold. 

In Ref.~\cite{Anderson:2013xka} a powerful algorithm for systematically generating all line bundle models $(k_a^i)$ with entries in the range $|k_a^i|\leq k_{\rm max}$ for a given upper bound $k_{\rm max}$ and selecting the models which satisfy all the above constraints has been outlined. In Ref.~\cite{Anderson:2013xka} this algorithm has been applied to the tetra-quadric, among other manifolds, and physically viable models have been extracted. The results are summarized in Table~\ref{tqmodels}. Altogether, $94$ viable GUT models for the available symmetry orders, $|\Gamma|=2,4,8,16$, are found. All these models correspond to consistent, anomaly-free and supersymmetric $SU(5)\times S(U(1)^5)$ GUT theories which satisfy the three-family constraint, have no $\overline{\bf 10}$ anti-families and at least one ${\bf 5}-\overline{\bf 5}$ pair to account for the Higgs doublets. Upon taking the quotient by $\Gamma$ and including a Wilson line, many of these become models with an MSSM spectrum and we will study one specific such example below. The complete dataset of viable line bundle sums can be accessed here \cite{lbdatabase}. The computational evidence that these models indeed represent the complete set of viable models on the tetra-quadric is presented in Fig.~\ref{SaturationPlots}, where the number of viable models with $|k_a^i|\leq k_{\rm max}$ is shown as a function of $k_{\rm max}$. For all symmetry orders, the number of models saturates at a value of $k_{\rm max}$ below $10$ and remains stable from thereon. In the next section, we will present various approaches to prove finiteness analytically. 
\begin{figure}[h!]
\begin{flushright}
$${\includegraphics[width=3.1in]{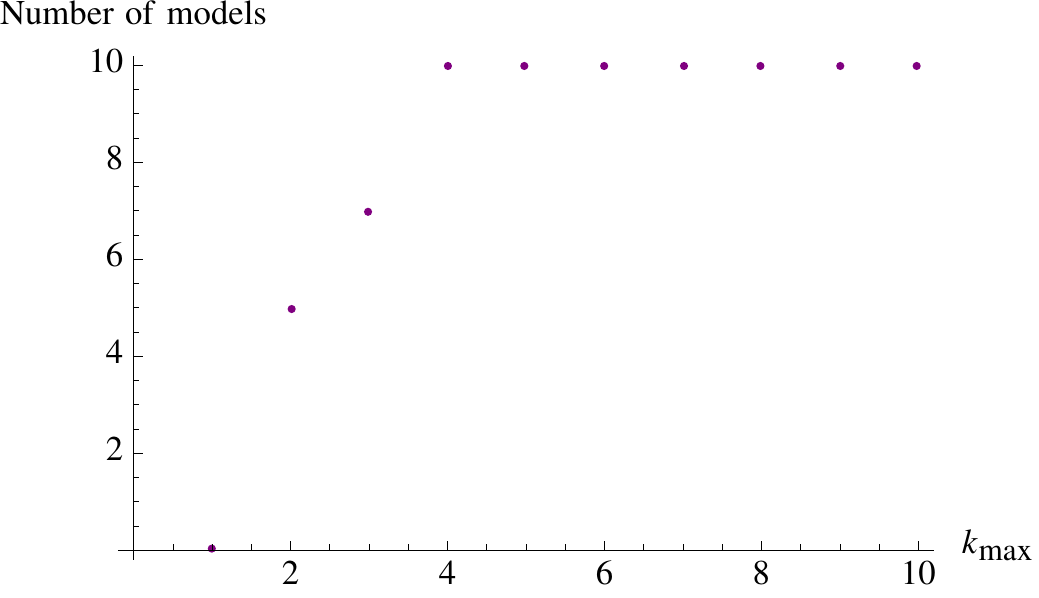}}
\hskip 2pt \lower 0pt\hbox{\includegraphics[width=3.1in]{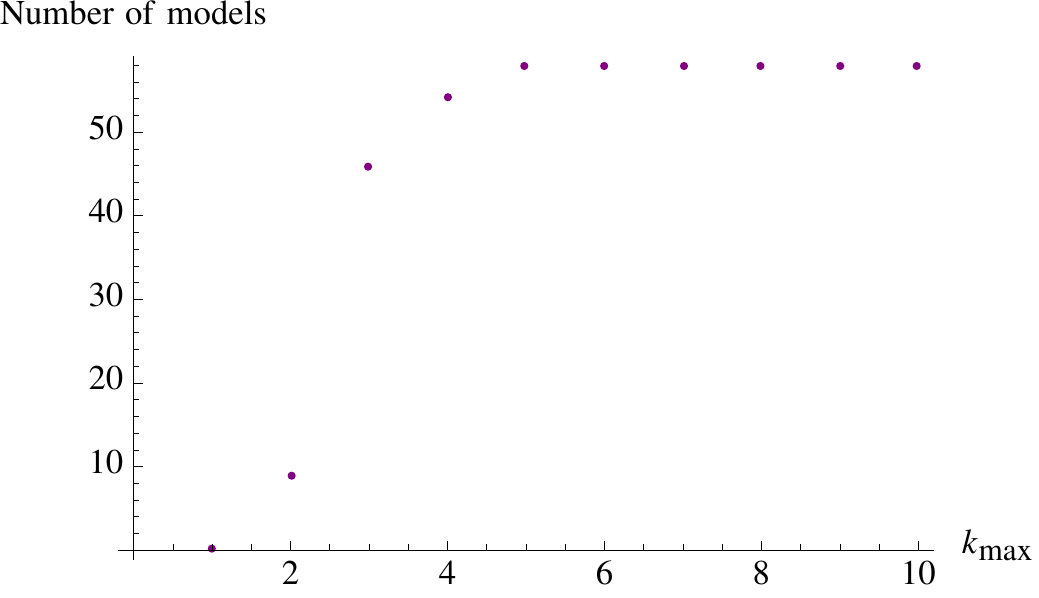}}
$$
\end{flushright}
\begin{flushright}
$${\includegraphics[width=3.1in]{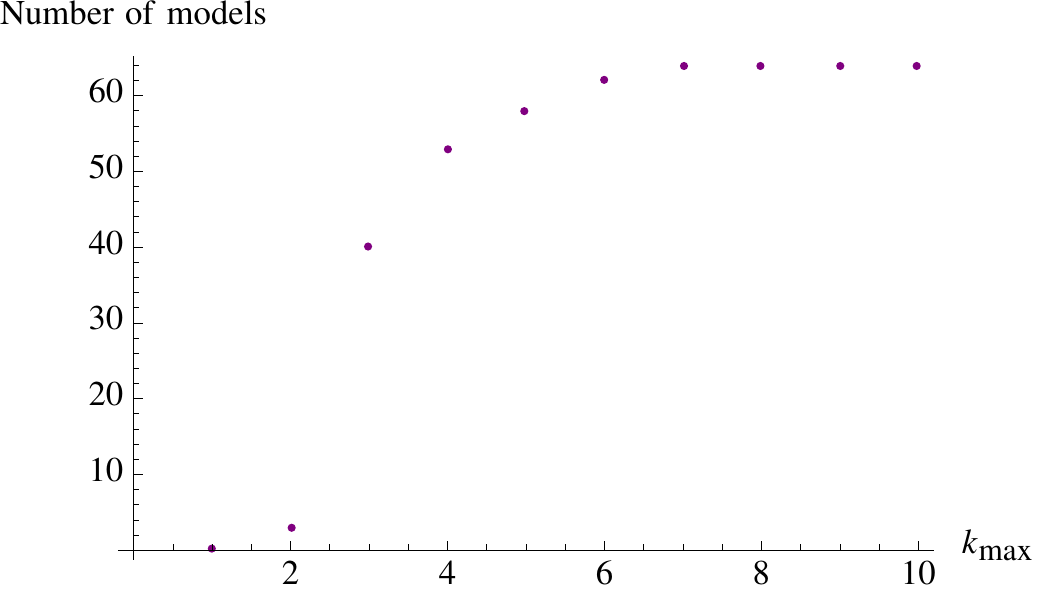}}
\hskip 2pt \lower 0pt\hbox{\includegraphics[width=3.1in]{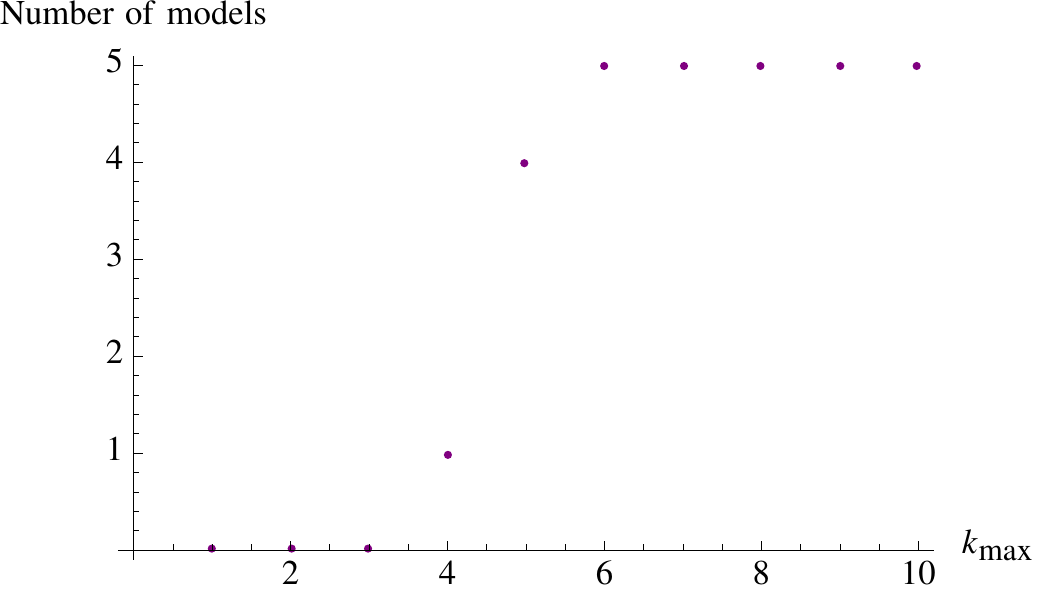}}
$$
\end{flushright}
\begin{center}
 \captionsetup{width=16cm}
\caption{\itshape The plots show the number of line bundle models (before imposing the absence of $\overline{\mathbf{10}}$ multiplets and the existence of $\mathbf{5}-\overline{\mathbf{5}}$ pairs) on the tetraquadric manifold as a function of the maximal line bundle entry in modulus. The four plots correspond, in order from top-left to bottom-right, to $|\Gamma| = 2, 4, 8$ and $16$.}\label{SaturationPlots}
\end{center}
\end{figure}
}

{\setstretch{1.25}
\section{Finiteness of the Class of Line Bundle Models}\label{sec:finiteness}
In this section we would like to discuss the problem of finiteness of poly-stable line bundle sums with fixed total Chern class analytically. Specifically, we would like to consider, for the example of the tetra-quadric manifold, the following claim.\\[0.3cm]
{\bf Claim: } On a given Calabi-Yau three-fold $X$ and for a fixed rank $n$, the number of line bundle sums
\begin{equation}
 V=\bigoplus_{a=1}^nL_a\; ,\quad L_a={\cal O}_X({\bf k}_a)\; , \label{lbsn}
 \end{equation}
 satisfying the following properties is finite:
\begin{itemize}
\item $c_1(V)=0$ or, equivalently, $\sum_{a=1}^n{\bf k}_a=0$.
\item The second Chern classes is constrained by $c_2(TX)-c_2(V)\in $ Mori cone of $X$.
\item All line bundles have vanishing slope, $\mu(L_a)=0$, simultaneously somewhere in the interior of the K\"ahler cone of $X$.
\item ${\rm ind}(V)=C$, where $C$ is a constant (here taken to be $C=-3|\Gamma|$, where $|\Gamma|$ is the order of a freely-acting symmetry $\Gamma$ on $X$).
\end{itemize}
The automated scan reported on in Ref.~\cite{Anderson:2013xka} and, for the case of the tetra-quadric, illustrated in Fig.~\ref{SaturationPlots} provides convincing evidence for the validity of this claim. In fact, as can be seen from those plots, all viable line bundle sums on the tetra-quadric satisfy $|k_a^i|\leq 10$. 

Unfortunately, analysing in a straightforward manner the way in which the various above constraints conspire to produce a finite class is untidy. For the tetra-quadric we carry this out explicitly in Appendix~\ref{app:finiteness}. In the next sub-section, we will prove the above claim for the tetra-quadric and the special case of rank two line bundle sums ($n=2$).  In Section~\ref{sec:physbound} we propose a transparent argument to derive a bound for line bundle sums of arbitrary rank; however, in order to make this argument feasible, we have to restrict the K\"ahler cone by imposing two constraints motivated from physics. We require that all K\"ahler moduli $t^i$ satisfy $t^i>1$, a constraint linked to the validity of the supergravity approximation, and finiteness of the Calabi-Yau volume, linked to the finiteness of low-energy coupling constants. 

\subsection{A workable example: rank two line bundle sums}\label{sec:tqbound}
Rank two line bundle sums with vanishing first Chern class have the form $V = L\oplus L^{^*}$, where $L={\cal O}_X({\bf k})$, so they are described by a single integer vector ${\bf k}$ with four entries $k^i$. We would like to show that the above claim is valid for this specific class of line bundle sums on the tetra-quadric. We begin with the slope zero condition which represents the main technical difficulty and recall that it can be written as
\begin{equation}
 \mu(L)=\kappa_i\,k^i\stackrel{!}{=}0\; ,
\end{equation}
where the quantities $\kappa_i=d_{ijk}\,t^j\,t^k$ have been explicitly given in Eq.~\eqref{kappai}. Here, the moduli $t^i$ have to be in the (interior of the) K\"ahler cone, $C_{\bf t}$, defined in Eq.~\ref{kc}. In order to avoid having to solve a quadric in $t^i$ we introduce the new coordinates ${\bf s}\in \IR^4$, defined by ${\bf s}=f({\bf t}) = \left( \kappa_1, \kappa_2, \kappa_3, \kappa_4\right) /4$. It turns out, and it is explicitly shown in Appendix~\ref{app:KahlerCone}, that the K\"ahler cone $f(C_{\bf t})$ in these new variables is a dense subset of the cone
\beq \label{kcs}
C_{{\bf s}} = \left\{ {\bf s}\in\IR^4 \ \left|\  {\bf n}_i\cdot {\bf s} \geq 0,\ {\bf e}_i\cdot {\bf s} \geq 0,\ 1\leq i\leq 4\right.\right\}\; .
\eeq
with ${\bf n}_i={\bf n}-{\bf e}_i$, ${\bf n}=(1,1,1,1)/2$ and the standard unit vectors ${\bf e}_i$ on $\mathbb{R}^4$. 
It is also useful to introduce the cone $\check{C}_{\mathbf k}$, dual to $C_{\mathbf s}$ which is given by the standard definition $\check{C}_{\mathbf k} = \{\, {\mathbf k} \in \IR^4 \left | {\mathbf k\cdot \mathbf s} \geq 0, \forall \mathbf s\in C_{\mathbf s} \right. \}$. A straightforward computation shows that it can also be written as
\beq
\check{C}_{\mathbf k} = \{\, {\mathbf k} \in \IR^4 \left | {\mathbf k\cdot \mathbf e}_{ij} \geq 0, \forall i<j \right. \}\; ,
\eeq
where ${\mathbf e}_{ij} = {\mathbf e}_i + {\mathbf e}_j$. We can now say that the slope zero condition is satisfied iff the equation ${\bf s}\cdot{\bf k}=0$ has a non-trivial solution in the interior $\mathring{C}_{\mathbf s}$ of the cone ${C}_{\mathbf s}$ and this, in turn, is equivalent to the condition $\mathbf k \notin \check{C}_{\mathbf k} \cup \left( -\check{C}_{\mathbf k}\right)$. Given the structure of the cone $\check{C}_{\mathbf k}$, this means the slope zero condition can be satisfied somewhere in the interior of the K\"ahler cone precisely if the vector ${\bf k}$ has two components $k_i, k_j$ with $k_i + k_j> 0$ and two components $k_l, k_m$ with $k_l + k_m<0$. Thus, up to permutations of the components of ${\bf k}$, the slope zero condition can be satsified iff
\beq\label{ineq1}
\begin{aligned}
&k_1 + k_2 > 0 \text{ and } k_1 + k_3 < 0 \text{ or } \\
&k_1 + k_2 > 0 \text{ and } k_3 + k_4 < 0
\end{aligned}
\eeq
Further, the bound on the second Chern class becomes
\begin{eqnarray}
 c_2^i(V)&=&4(k^2k^3+k^2k^4+k^3k^4,k^1k^3+k^1k^4+k^3k^4,k^1k^2+k^1k^4+k^2k^4,k^1k^2+k^1k^3+k^2k^3)\nonumber\\
 &\stackrel{!}{\leq}&(24,24,24,24)\; .\label{ineq2}
\end{eqnarray}
It can be shown, for example using Mathematica, that the system of integer inequalities given by Eqs.~(\ref{ineq1}) and (\ref{ineq2}) has a finite number of solutions $\mathbf k = (k_1,k_2,k_3,k_4)$ which all satisfy $-7< k_i <7$. We note that, in arriving at this result, we have not even used the constraint on the index of $V$. 

\subsection{A bound rooted in Physics}\label{sec:physbound}
Let us now present a more general finiteness proof which applies to line bundle sums of any rank on the tetra-quadric (and can indeed be applied to other Calabi-Yau manifolds) which, however, requires two additional, physically motivated assumptions.

First recall that the K\"ahler moduli space metric \cite{Candelas:1990pi} for a Calabi-Yau manifold can be written as
\begin{equation}
G_{ij} = \frac{1}{2\,\text{Vol}(X)} \int_X J_i\wedge \star J_j = -3 \left( \frac{\kappa_{ik}}{\kappa} - \frac{2\kappa_i\kappa_j}{3\kappa^2}\right) 
\end{equation}
where $\text{Vol}(X)=\kappa/6$ is the Calabi-Yau volume with respect to the Ricci-flat metric, $\kappa = d_{ijk}\, t^i\,t^j\,t^k$, $\kappa_i = d_{ijk}\,t^j\,t^k$ and $\kappa_{ij}=d_{ijk}\,t^k$. The slope zero conditions for a line bundle sum of the form~\eqref{lbsn} can be written as
\begin{equation}
 \kappa_ik_a^i=0\; .
\end{equation} 
Now consider the sum 
\begin{equation}\label{eq:bound1}
\sum_a {\bf k}_a^T G\, {\bf k}_a = -\frac{3}{\kappa} d_{ijk} \sum_a k_a^i\, k_a^j\, t^k = -\frac{6}{\kappa} t^i\, \text{ch}_{2i}(V) \leq \frac{6}{\kappa} |{\bf t}| |\text{ch}_{2i}(V) |\leq \frac{6}{\kappa}|{\bf t}||c_{2i}(TX)|\; .
\end{equation}
Introducing the modified moduli space metric $\widetilde G  = \kappa\, G /( 6 |{\bf t}|)$ this means that
\begin{equation}
 \sum_a {\bf k}_a^T \widetilde{G}\, {\bf k}_a\leq |c_{2i}(TX)|\; . \label{kcons}
 \vspace{-9pt}
\end{equation} 
For a fixed K\"ahler class ${\bf t}$ in the interior of the K\"ahler cone, the moduli space metric $G$ and indeed $\widetilde{G}$ are positive definite and, hence, the inequality~\eqref{kcons} constrains the available integer vectors ${\bf k}_a$ to a finite set. This statement applies to all Calabi-Yau three-folds. However, it has a limitation which is relevant for the physics application we are discussing. In physics, we are not interested in fixing the K\"ahler class, that is, different line bundle sums can satisfy the slope zero conditions for different loci in K\"ahler moduli space. In particular, we cannot, by the above argument, exclude a sequence of line bundle sums whose associated slope zero loci approach the boundary of the K\"ahler cone. In such a situation, the eigenvalues of $\widetilde G$ are no longer bounded from below and the above finiteness argument breaks down.

One way to resolve this difficulty is to restrict the ``allowed" region in K\"ahler moduli space, that is, in essence, exclude points close to the boundary. Specifically, what we require is that all $t^i>1$ (assuming the K\"ahler cone is given by $t^i\geq 0$, as is the case for the tetra-quadric) and that $\text{Vol}(X)\lesssim V_{\rm max}$, for a maximal volume $V_{\rm max}$. We are then asking about the number of line bundle sums satisfying all conditions listed in the above claim plus the additional requirement that the slope zero conditions hold in the so-defined portion of K\"ahler moduli space. The physical motivations for these two conditions are the validity of the supergravity approximation (which requires the internal space to be larger than one in string units) and the finiteness of the low-energy coupling constants (specifically finiteness of the gauge couplings and Newton's constant which are related to the volume).
\begin{figure}[h!]
\vspace{12pt}
\begin{center}
\includegraphics[width=4in]{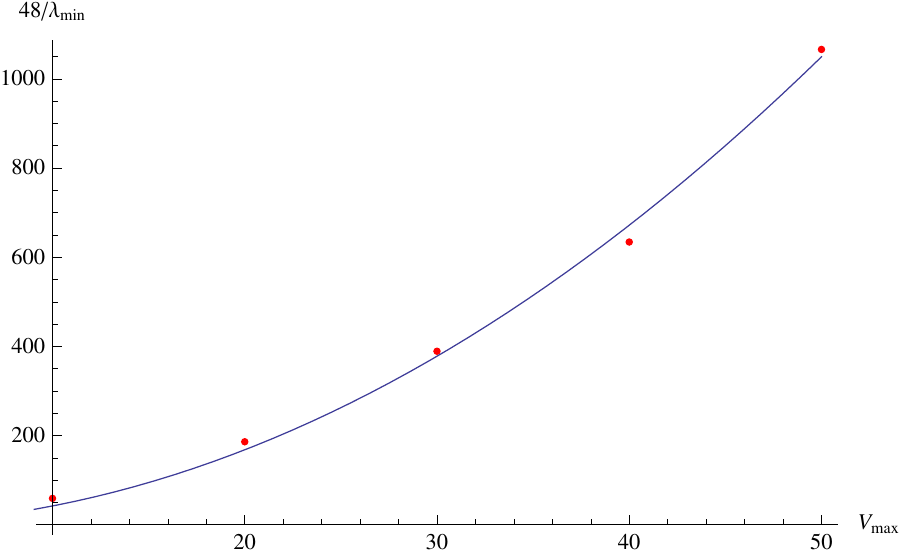}
\captionsetup{width=16cm}
\caption{\itshape The plot shows the dependence of the bound $48/\lambda_{\text{min}}$ (vertical axis) on the Calabi-Yau volume (horizontal axis). The red points represent the values obtained by numerical methods. The blue curve represents the best fit with a parabola passing through the origin. }\label{BoundVsVolume2}
\end{center}
\end{figure}
This method can be applied to any Calabi-Yau manifold, for which the eigenvalues of $\widetilde{G}$ are bounded from below over the region of K\"ahler moduli space defined above. Let us denote by $\lambda_{\rm min}$ the minimum eigenvalue assumed by $\widetilde{G}$ over the specified portion of K\"ahler moduli space. Clearly, the value of $\lambda_{\rm min}$ depends on the specific Calabi-Yau manifold and the maximal value, $V_{\rm max}$, of the volume. If indeed $\lambda_{\rm min}>0$ then Eq.~\eqref{kcons} leads to the bound
\begin{equation}
 \sum_a |{\bf k}_a|^2\leq\frac{|c_2(TX)|}{\lambda_{\rm min}}\; . \label{kbound1}
\end{equation} 

Let us carry this out explicitly for the tetra-quadric. In this case, the metric $\widetilde{G}$ is explicitly given by
\beq
\widetilde G_{ij} = \frac{1}{\sqrt{t_1^2+t_2^2+t_3^2+t_4^2}}\, \left({\displaystyle \sum_{a<b<c} t_a\,t_b\,t_c}\right)^{-1}\, {\displaystyle \sum_{\stackrel{a<b}{a,b\neq i,j}} t_a^2\, t_b^2}
\eeq
and the bound~\eqref{kbound1} specializes to
\begin{equation}
 \sum_a |{\bf k}_a|^2\leq\frac{48}{\lambda_{\rm min}}\; . \label{kbound2}
\end{equation}
The value of $48/\lambda_{\text{min}}$ as a function of $V_{\rm max}$, determined by a straightforward numerical scan over the relevant portion of the moduli space, has been plotted in Figure~\ref{BoundVsVolume2}. For the explicit models in Table~\ref{tqmodels} whose slope zero locus intersects the region defined by $t^i>1$ and ${\rm Vol}(X)\leq 50$ we find that $\sum_a|{\bf k}_a|^2<50$. Comparing with Fig.~\ref{BoundVsVolume2} this means that, while the bound~\eqref{kbound2} holds, it is actually rather weak and presumably of limited practical use.
}

{\setstretch{1.15}
\section{A line bundle model on the tetra-quadric}\label{sec:model}
In this chapter, we present a specific example taken from the set of phenomenologically viable line bundle models on the tetra-quadric described in the previous section. It is in the context of this model that we will study the question of continuation into the non-Abelian part of the bundle moduli space and the implications for the mass of the Higgs doublets. 
\subsection{Definition of the model}
The bundle $V$ for the model in question is given by the sum of the following five line bundles
\begin{equation}\label{lbs}
 \begin{array}{lllllllllll}
 L_1&=&{\cal O}_X(-1,0,0,1)&,&L_2&=&{\cal O}_X(-1,-3,2,2)&,&L_3&=&{\cal O}_X(0,1,-1,0)\\
 L_4&=&{\cal O}_X(1,1,-1,-1)&,&L_5&=&{\cal O}_X(1,1,0,-2)
 \end{array}
\end{equation}
so the associated matrix $(k_a^i)$ of line bundle integers reads
\begin{equation}\label{eq:example}
(k_a^i)=
\cicy{ \\ \\ \\ \\ }
{ -1 & -1 & ~~0 & ~~1 & ~~1~ \\
~~0 & -3 & ~~1 & ~~1 & ~~1~ \\
~~0 & ~~2 & -1 & -1 & ~~0 ~\\
 ~~1 & ~~2 &~~ 0 & -1 & -2 ~\\}\; .
\end{equation}
The rows of this matrix sum up to zero, so clearly we have $c_1(V)=0$, as required. From Eq.~\eqref{c2} we find
\begin{equation}
 c_{2i}(V)=(24,8,20,12)
\end{equation} 
so that the anomaly constraint~\eqref{tqanom} is satisfied. Further, with ${\rm rank}(k_a^i)=3$, the rank constraint is satisfied and all line bundle slopes~\eqref{tqslope0} are zero on the ray in K\"ahler moduli space where $\kappa_1=\kappa_2=\kappa_3=\kappa_4$ which corresponds to the diagonal $t_1=t_2=t_3=t_4$. Altogether this means we have defined a consistent, supersymmetric GUT model with symmetry $SU(5)\times S(U(1)^5)$. Since ${\rm rank}(k_a^i)=3$ one linear combination of the $U(1)$ symmetries is non-anomalous with a massless vector boson at the Abelian locus. This specific linear combination is $(0,1,2,0,1)$, the non-trivial vector  in the kernel of the matrix $(k_a^i)$.

\subsection{The GUT spectrum at the Abelian locus}
The total dimensions of the relevant cohomology groups (computed, e.g.~using the formulae of Appendix~\ref{app:tqcoh}) are given by
\beq\label{cohV}
\begin{aligned} 
h^{^{\!\bullet}}(X,V)\ \, & =\  (0,12,0,0) \\
h^{^{\!\bullet}} (X,\wedge^2V)&=\ (0, 15, 3, 0)\\
h^{^{\!\bullet}} (X,V\otimes V^{^*})&=\ (5, 60, 60, 5)
\end{aligned}
\eeq
Hence, we have a total of $12$ GUT families in ${\bf 10}\oplus\overline{\bf 5}$ plus three ${\bf 5}-\overline{\bf 5}$ pairs and a large number of singlet fields. In order to quotient this GUT model to a three-family standard model we need $|\Gamma|=4$ and with $\Gamma=\mathbb{Z}_2\times\mathbb{Z}_2$ we have such a symmetry available on the tetra-quadric. Before we discuss this in detail we should be more precise on how the GUT spectrum is split up into the various line bundle sectors. For this we compute the following relevant line bundle cohomologies

\begin{equation}
\begin{array}{lllllll}
 h^{^{\!\bullet}}(X,L_2)&=&(0,8,0,0)&,&h^{^{\!\bullet}}(X,L_5)&=&(0,4,0,0)\\[4pt]
 h^{^{\!\bullet}}(X,L_2\otimes L_4)&=&(0,4,0,0)&,&h^{^{\!\bullet}}(X,L_2\otimes L_5)&=&(0,3,3,0)\\[4pt]
 h^{^{\!\bullet}}(X,L_4\otimes L_5)&=&(0,8,0,0)&,&h^{^{\!\bullet}}(X,L_1\otimes L_2^*)&=&(0,0,12,0)\\[4pt]
 h^{^{\!\bullet}}(X,L_1\otimes L_5^*)&=&(0,0,12,0)&,&h^{^{\!\bullet}}(X,L_2\otimes L_3^*)&=&(0,20,0,0)\\[4pt]
 h^{^{\!\bullet}}(X,L_2\otimes L_4^*)&=&(0,12,0,0)&,&h^{^{\!\bullet}}(X,L_3\otimes L_5^*)&=&(0,0,4,0)
\end{array} 
\vspace{12pt}
\end{equation}
Here, we have dropped all entirely zero cohomologies. This gives rise to the following spectrum
\begin{equation}
 8\, {\bf 10}_2\,,\; 4\,{\bf 10}_5\,,\;4\,\overline{\bf 5}_{2,4}\,,\;3\,\overline{\bf 5}_{2,5}\,,\;8\,\overline{\bf 5}_{4,5}\,,\; 3{\bf 5}_{2,5}\,,\;
 12\,{\bf 1}_{2,1}\,,\;12\,{\bf 1}_{5,1}\,,\;20\,{\bf 1}_{2,3}\,,\;12\,{\bf 1}_{2,4}\,,\;4\,{\bf 1}_{5,3}\; .\label{gutspec}
\end{equation}

\subsection{The Standard Model spectrum at the Abelian locus}
The relevant $\Gamma=\mathbb{Z}_2\times\mathbb{Z}_2$ symmetry for the above model is the one whose generators are given by the action of the matrices
\vspace{12pt}
\begin{equation}\label{Z2Z2def}
 \left(\begin{array}{rr}1&0\\0&-1\end{array}\right)\; ,\quad \left(\begin{array}{rr}0&1\\1&0\end{array}\right)
\vspace{12pt}
\end{equation}
simultaneously on the coordinates of all four $\IC\IP^1$ ambient space factors. For an appropriate choice of equivariant structure and Wilson line, forming the quotient by this symmetry leads to the downstairs spectrum~\footnote{Our notation is slightly judicious in that, strictly, we cannot decide at this stage which linear combination of the four available doublets is the down Higgs $H$ and which ones are the three lepton doublets.}
\begin{equation}
 2\, {\bf 10}_2\,,\; {\bf 10}_5\,,\;\overline{\bf 5}_{2,4}\,,\;2\,\overline{\bf 5}_{4,5}\,,\; H_{2,5}\,,\;\overline{H}_{2,5}\,,\;
 3\,{\bf 1}_{2,1}\,,\;3\,{\bf 1}_{5,1}\,,\;5{\bf 1}_{2,3}\,,\;3\,{\bf 1}_{2,4}\,,\;{\bf 1}_{5,3}\; . \label{smspec}
\end{equation} 
Hence, we have precisely three standard model families (which we have listed in GUT notation but should be thought of as being broken up as ${\bf 10}_a\rightarrow (Q_a,u_a,e_a)$ and $\overline{\bf 5}_{a,b}\rightarrow (d_{a,b},L_{a,b})$ into standard model multiplets), one pair of Higgs doublets and $15$ bundle moduli singlets. We note that the $U(1)$ charges are the same for all standard model multiplets originating from the same GUT multiplet and, hence, for the purpose of discussing the implications of $S\left(U(1)^5\right)$ invariance, keeping the GUT notation is adequate. 

\subsection{The superpotential}\label{sec:W}
The superpotential for the fields~\eqref{gutspec} is highly constrained by the $S(U(1)^5)$ charges. At the GUT level the only allowed terms, including operators with singlet insertions are
\begin{equation}
 W=\lambda_{IJK} {\bf 5}_{2,5}^{(I)}{\bf 10}_2^{(J)}{\bf 10}_5^{(K)}+\rho_{IJK} {\bf 1}_{2,4}^{(I)}\overline{\bf 5}_{4,5}^{(J)}{\bf 5}_{2,5}^{(K)}\; ,
 \label{Wgut}
\end{equation} 
where the indices $I,J,K\ldots $ run over various ranges, as indicated by the multiplicities in the GUT spectrum~\eqref{gutspec} and $\lambda_{IJK}$ and $\rho_{IJK}$ are arbitrary couplings. At the standard model level, the analogous terms for the spectrum~\eqref{smspec} are
\begin{equation}
 W=\lambda_i\overline{H}_{2,5}(Q_2^{(i)}u_5+Q_5u_2^{(i)})+\rho_{\alpha i} {\bf 1}_{2,4}^{(\alpha)}L_{4,5}^{(i)}\overline{H}_{2,5}\; , \label{Wsm}
\end{equation} 
where $i=1,2$ labels the two ${\bf 10}_2$ families and the two lepton doublets $L_{4,5}$ from the two $\overline{\bf 5}_{4,5}$ multiplets and $\alpha=1,2,3$ labels the three singlets ${\bf 1}_{2,4}$.

These results have a number of important implications for the structure of the model and its phenomenology. To discuss this, let us focus on the standard model superpotential~\eqref{Wsm} for concreteness, although analogous statements follow for its GUT counterpart~\eqref{Wgut}. The presence of the Yukawa terms means that the up quark mass matrix has rank two and, while a rank one matrix may be preferably at this level, this means a perturbative and generically large top Yukawa coupling is present. The down quark and lepton Yukawa matrices are entirely zero at the perturbative level so, for a realistic model, they would have to be generated non-perturbatively. Further, all operators at dimension four and five which can lead to fast proton decay are forbidden. The point is that, while we certainly do not advertise this model as the one and only standard model from string theory, it does have modestly attractive phenomenological properties and provides a semi-realistic setting for the analysis of the bundle moduli space which we will carry out in the remaining part of the paper.

Specifically, our intention is to explore the moduli space of non-Abelian bundles for which the line bundle sum~\eqref{eq:example} arises as a special locus. From the viewpoint of the four-dimensional effective field theory, the Abelian locus is characterized by the vanishing VEVs of all singlet fields ${\bf 1}_{a,b}$, while switching on such VEVs corresponds to moving away from the Abelian locus into the non-Abelian part of the moduli space. From this point of view, the last term in the superpotential~\eqref{Wsm} for our example model is the most interesting one. At the Abelian locus where, in particular, $\langle {\bf 1}_{2,4}\rangle=0$ this term is simply a coupling. However, for $\langle {\bf 1}_{2,4}\rangle\neq 0$ this term will lead to a mass for the Higgs doublets (or rather for the up Higgs and one linear combination of what we have called lepton doublets) and essentially remove the Higgs from the low-energy spectrum. On the other hand, the spectrum~\eqref{smspec} contains many other singlets which do not appear in the superpotential. A continuation into the non-Abelian part of the moduli space along those singlet directions, while keeping $\langle {\bf 1}_{2,4}\rangle=0$, should leave the Higgs doublets massless. Phrased in terms of the GUT theory, the structure of the superpotential~\eqref{Wgut} suggests that three ${\bf 5}-\overline{\bf 5}$ pairs are removed from the low-energy spectrum if $\langle {\bf 1}_{2,4}\rangle\neq 0$ but that these states remain massless in all parts of the non-Abelian moduli space where $\langle {\bf 1}_{2,4}\rangle=0$. One goal for the remainder of the paper is to verify these statements from a more fundamental viewpoint, that is, by explicitly constructing families of non-Abelian bundles and computing their cohomology. 
}

{\setstretch{1.15}
\section{Non-Abelian deformations}\label{sec:monads}
In this section, we would like to discuss two ways of constructing non-Abelian bundles which split into a given line-bundle sum at a specific locus in moduli space and apply these methods to the example presented in the previous section. We will focus on two main bundle constructions, namely extensions and monads.

\subsection{Extensions of line bundle sums}
Extension bundles provide a method of constructing non-Abelian bundles which split into a given set of line bundles
\begin{equation}
 V=\bigoplus_{a=1}^nL_a\; .
\end{equation} 
Group the line bundles into two sets, indexed by $I\subset \{1,\ldots ,n\}$ and $\bar{I}=\{1,\ldots ,n\}\backslash I$, define the sub-bundles $V_I=\bigoplus_{a\in I}L_a$ and $V_{\bar{I}}=\bigoplus_{a\in\bar{I}}L_a$, and write down the extension sequence
\begin{equation}
 0\ \longrightarrow\ V_I\ \longrightarrow\ \widetilde{V}\ \longrightarrow\ V_{\bar{I}}\ \longrightarrow\ 0\; . \label{extseq}
\end{equation} 
The moduli space of the extension bundles $\widetilde{V}$ defined by this short exact sequence is given by
\begin{equation}
 {\rm Ext}^1(V_{\bar{I}},V_I)\cong H^1(X,V_I\otimes V_{\bar{I}}^{^*})=\bigoplus_{a\in I, b\in{\bar I}}H^1(X,L_a\otimes L_b^{^*})\; . \label{ext}
\end{equation} 
The origin of this space corresponds to the split bundle $\widetilde{V}=V_I\oplus V_{\bar{I}}=V$, but away from it $\widetilde{V}$ becomes non-Abelian. Note that the tangent space to the bundle moduli space of $V$ is given by
\begin{equation}
 H^1(X,V^{^*}\!\otimes V)=\bigoplus_{a\neq b}H^1(X,L_a\otimes L_b^{^*})
\end{equation} 
and, in general, this is larger than~\eqref{ext}. However, if $H^1(X,L_a\otimes L_b^{^*})\neq 0$ only if $a\in I$ and $b\in \bar{I}$ for a suitable choice of $I$ and $\bar{I}$ then the two spaces are indeed the same and the extension sequence captures the full set of non-Abelian deformations. We note that the cohomology $H^1(X,L_a\otimes L_b^{^*})$ contains the low-energy singlets earlier denoted ${\bf 1}_{a,b}$. So, if the $U(1)$ charges $\{1,\ldots ,n\}$ can be split into two disjoint subsets $I$ and $\bar{I}$ such that only singlets ${\bf 1}_{a,b}$ with $a\in I$ and $b\in{\bar I}$ exist then the extension sequence~\eqref{extseq} is ``complete".

We recall from Eq.~\eqref{gutspec} that the singlet spectrum for our tetra-quadric example consists of ${\bf 1}_{2,1}$, ${\bf 1}_{5,1}$, ${\bf 1}_{2,3}$, ${\bf 1}_{2,4}$, ${\bf 1}_{5,3}$. Hence, for $I=\{2,5\}$ and $\bar{I}=\{1,3,4\}$ the above completeness condition is indeed satisfied and the relevant extension sequence for our example reads
\begin{equation}
 0\ \longrightarrow\ V_I\ \longrightarrow\ \widetilde{V}\ \longrightarrow V_{\bar{I}}\ \longrightarrow\ 0\; ,\quad V_I=L_2\oplus L_5\; ,\quad V_{\bar{I}}=L_1\oplus L_3\oplus L_4\label{extV}
\end{equation} 
The line bundles $L_1,\dots ,L_5$ have been defined in Eq.~\eqref{lbs}.

We would now like to compute the relevant cohomologies of the so-defined extension bundle $\widetilde{V}$ and compare these with the cohomologies of the line bundle sum $V$. For $\widetilde{V}$ we can simply consider the long exact sequence associated to the extension sequence~\eqref{extV}. With $h^{^{\!\bullet}}(V_I)=(0,12,0,0)$ and $h^{^{\!\bullet}}(V_{\bar{I}})=(0,0,0,0)$ this long exact sequence reads
\begin{equation}
 \begin{array}{llllll}
 &V_I&\longrightarrow& \widetilde{V}&\longrightarrow& V_{\bar{I}}\\[8pt]
 h^0(X,\cdot\,)$\ \ \ \ $&0&&0&&0\\[3pt]
 h^1(X,\cdot\,)$\ \ \ \ $&12&&12&&0\\[3pt]
 h^2(X,\cdot\,)$\ \ \ \ $&0&&0&&0\\[3pt]
 h^3(X,\cdot\,)$\ \ \ \ $&0&&0&&0
\end{array}\; 
\end{equation} 
 so that $h^{^{\!\bullet}}(\widetilde{V})=(0,12,0,0)$. This coincides with the result for the cohomology of $V$ in Eq.~\eqref{cohV}. Since the index is unchanged, that is ${\rm ind}(\widetilde{V})={\rm ind}(V)=12$, (and we do not expect new vector-like states to appear in the non-Abelian region) this result is entirely expected. Physically, it means that all the ${\bf 10}$ multiplets which arise at the Abelian locus remain massless when moving into the non-Abelian part of the moduli space. 

The situation is considerably more complicated for the ${\bf 5}$ and $\overline{\bf 5}$ multiplets which arise from the cohomology of $\wedge^2\widetilde{V}$. As before the index is unchanged,  ${\rm ind}(\wedge^2\widetilde{V})={\rm ind}(\wedge^2V)=12$, so that the twelve $\overline{\bf 5}$ multiplets remain massless. However, the fate of the three vector-like ${\bf 5}-\overline{\bf 5}$ multiplets and, hence, the fate of the Higgs doublets is harder to decide. We begin with the second wedge power sequence
\begin{equation}
 0\ \longrightarrow\  \wedge^2V_I\ \longrightarrow\ \wedge^2\widetilde{V}\ \longrightarrow\ \widetilde{V}\otimes V_{\bar{I}}\ \longrightarrow\ S^2V_{\bar{I}}\ \longrightarrow\ 0\; ,
 \label{W2}
\end{equation}
associated to the extension sequence~\eqref{extV}. To determine the required cohomology of $\widetilde{V}\otimes V_{\bar{I}}$ we tensor the extension sequence~\eqref{extV} with $V_{\bar{I}}$ and, with $h^{^{\!\bullet}}(V_I\otimes V_{\bar{I}})=(0,12,0,0)$ and $h^{^{\!\bullet}}(V_{\bar{I}}\otimes V_{\bar{I}})=(0,6,6,0)$, this leads to
\begin{equation}
 \begin{array}{cccccc}
 &V_I\otimes V_{\bar{I}}&\longrightarrow& \widetilde{V}\otimes V_{\bar{I}}&\longrightarrow& V_{\bar{I}}\otimes V_{\bar{I}}\\[8pt]
 h^0(X,\cdot\,)$\ \ \ \ $&0&&0&&0\\[3pt]
 h^1(X,\cdot\,)$\ \ \ \ $&12&&18&&6\\[3pt]
 h^2(X,\cdot\,)$\ \ \ \ $&0&&6&&6\\[3pt]
 h^3(X,\cdot\,)$\ \ \ \ $&0&&0&&0
\end{array}\; ,
\end{equation} 
This means that $h^{^{\!\bullet}}(\widetilde{V}\otimes V_{\bar{I}})=(0,18,6,0)$ and, in terms of spaces, that
\begin{eqnarray}
 H^1(X,\widetilde{V}\otimes V_{\bar{I}})&\cong& H^1(X,V_I\otimes V_{\bar{I}})\oplus H^1(X,V_{\bar{I}}\otimes V_{\bar{I}})\label{coh1}\\[4pt]
 H^2(X,\widetilde{V}\otimes V_{\bar{I}})&\cong&H^2(X,V_{\bar{I}}\otimes V_{\bar{I}})\label{coh2}\; .
\end{eqnarray} 
 We can now split the wedge power sequence~\eqref{W2} into two short exact sequences
 \begin{equation}\label{W2split}
  \begin{array}{cccccccccccccc}
 &\wedge^2V_I&\longrightarrow& \wedge^2\widetilde{V}&\longrightarrow& K&\quad\quad\quad\quad&K&\longrightarrow&\widetilde{V}\otimes V_{\bar{I}}&\longrightarrow&S^2V_{\bar{I}}\\[8pt]
 h^0(X,\cdot\,)$\ \ \ \ $&0&&0&&0& &0&&0&&0\\[3pt]
 h^1(X,\cdot\,)$\ \ \ \ $&3&&12+c&&12& &12&&18&&6\\[3pt]
 h^2(X,\cdot\,)$\ \ \ \ $&3&&c&&0& &0&&6&&6\\[3pt]
 h^3(X,\cdot\,)$\ \ \ \ $&0&&0&&0& &0&&0&&0
\end{array}
\end{equation} 
where $\delta:H^1(X,K)\rightarrow H^2(X,\wedge^2V_I)$ and $c={\rm dim}\, {\rm Coker}(\delta)$. The key is now to compute the Coker dimension $c$ of this map. For $c=3$ we have three massless ${\bf 5} - \overline{\bf 5}$ pairs while for $c=0$ they have disappeared from the low-energy spectrum. From Eqs.~\eqref{coh1}, \eqref{coh2}, the source and target space for the map $\delta$ can be written as
\begin{eqnarray}
 H^1(X,K)&\cong& H^1(X,V_I\otimes V_{\bar{I}})=\bigoplus_{a\in I,b\in\bar{I}}H^1(X,L_a\otimes L_b)\nonumber\\
 &=&H^1(X,L_2\otimes L_4)\oplus H^1(X,L_4\otimes L_5)\label{source}\\
 H^2(X,\wedge^2V_I)&\cong& H^2(L_2\otimes L_5)
 \end{eqnarray}
 while the map itself resides in
 \begin{equation}
 \delta\in H^1(X,(V_I\otimes V_{\bar{I}})^{^*}\!\otimes \wedge^2V_I)=\bigoplus_{a\in I,b\in\bar{I}}H^1(X,L_a^{^*}\!\otimes L_b^{^*}\!\otimes L_2\otimes L_5)\; . \label{map}
\end{equation} 
A comparison between Eqs.~\eqref{source} and \eqref{map} shows that only the components $(a,b)= (2,4),(5,4)$ of the map are relevant for the given source space. It turns out that the $(2,4)$ component of the map $H^1(L_5\otimes L_4^{^*})$ vanishes so that the non-trivial part of the map $\delta$ is characterized by
\begin{equation}
 \delta:H^1(X,L_4\otimes L_5)\rightarrow H^2(X,L_2\otimes L_5)\quad\mbox{ where }\quad \delta\in H^1(X,L_2\otimes L_4^{^*})\; .
\end{equation} 
The crucial observation is that the map only depends on $H^1(X,L_2\otimes L_4^{^*})$ which is precisely the cohomology containing the $12$ bundle moduli singlets earlier denoted by ${\bf 1}_{2,4}$. Hence, if the VEVs of those singlets vanish the map $\delta$ is trivial so that, independently of the other singlet field values, $c={\rm dim}\, {\rm Coker}(\delta)=3$. In this case, from \eqref{W2split}, the cohomology calculation leads to three massless ${\bf 5}-\overline{\bf 5}$ pairs, in accordance with the effective field theory expectation explained in Section~\ref{sec:W}. On the other hand, if the ${\bf 1}_{2,4}$ singlets have non-zero VEVs, the map $\delta$ becomes non-trivial and, as a result, $c<3$. For generic values of the ${\bf 1}_{2,4}$ VEVs the expectation is that $c=0$, so that $h^2(X,\wedge^2\widetilde{V})=3$ and all three ${\bf 5}-\overline{\bf 5}$ pairs are removed from the spectrum. Again, this conforms with the expectation from the four-dimensional effective field theory. 

We would now like to explore non-Abelian continuations of line bundle sums using a different method - the monad construction.
}

{\setstretch{1.15}
\subsection{Monads from line bundle sums}
Monad bundles provide a relatively straightforward way to construct bundles with non-Abelian structure groups from line bundle sums. On a Calabi-Yau manifold $X$, define two line bundle sums
\begin{equation}
 B = \bigoplus_{\alpha=1}^{\text{rk}(B)}{\cal O}_X({\bf b}_\alpha)\; ,\quad C = \bigoplus_{\mu=1}^{\text{rk}(C)} {\cal O}_X({\bf c}_\mu)
 \end{equation}
and the bundle $\widetilde{V}$ by the short exact sequence
\begin{equation} \label{eq:monad}
0\ \longrightarrow\ \widetilde{V} \longrightarrow\ B\ \stackrel{f}{\longrightarrow}\ C\ \longrightarrow\ 0  
\end{equation}
so that $\widetilde{V}\cong{\rm Ker}(f)$. Here, $f\in \text{Hom}(B,C)\cong H^0(X,C\otimes B^{^*})$ can be thought off as a ${\rm rk}(C)\times{\rm rk}(B)$ matrix with entries $f_{\mu\alpha}\in H^0(X,{\cal O}_X({\bf c}_\mu- {\bf b}_\alpha)$. Of course, it has to be checked that the above sequence is indeed exact for a given choice of map $f$ which amounts to verifying that ${\rm Im}(f)=C$. 

\subsubsection{Monads with split loci}\label{sec:monadconstruction}
Typically, a family of monad maps $f$, parametrized by the coefficients of the polynomials $f_{ai}$, is available leading to a family of monad bundles $\widetilde{V}$. Our task is to construct such a family of monad bundles which splits into a given line bundle sum $V$ at a particular locus in moduli space, that is, for a specific sub-class of maps $f$. 

An obvious way to proceed would be to define the line bundle sum $B$ in the monad sequence as $B = V\oplus \widetilde B$ where $V$ is the given line bundle sum and $\widetilde B$ is some other sum of line bundles. If the monad bundle $\widetilde{V}$ splits into $V$ at some locus, it follows that ${\rm ch}(\widetilde{V})={\rm ch}(V)$. This necessary property can be built into the construction by choosing the line bundle sum $C$ in the monad sequence such that  $\text{ch}(\widetilde{B}) =\text{ch}(C)$. The monad map then has a block diagonal structure $f=(g,h)$, where $g$ corresponds to the $V$-part of $B$ and $h$ corresponds to $\widetilde B$. At the locus in bundle moduli space where $g=0$, we have $V\subset \text{Ker}(f)$ and this should correspond to the desired split locus. The problem with this construction is that $\text{Ker}(f)$ may not be a vector bundle when $f=(0,h)$. Indeed, for the case of the tetra-quadric, the (quadratic) matrix $h$ degenerates on a co-dimension one locus $\text{det} (h)=0$ in the ambient space, which, generically, intersects the tetra-quadric hypersurface. 

\vspace{10pt}
To avoid this problem we will use the following alternative construction. We start with a monad realisation of the structure sheaf
\begin{equation}\label{eq:structure_sheaf}
0\ \longrightarrow\ \cO_X \longrightarrow\ \widetilde{B}_a\ \stackrel{f_a}{\longrightarrow}\ \widetilde{C}_a\ \longrightarrow\ 0  
\end{equation}
where $\widetilde{B}_a$ and $\widetilde{C}_a$ are sums of line bundles satisfying 
\begin{equation}
\text{rk} (\widetilde{B}_a) = \text{rk}(\widetilde{C}_a) +1 \ \ \ \text{and}\ \ \ c_1(\widetilde{B}_a) = c_1(\widetilde{C}_a)\; .
\end{equation}
On the tetra-quadric, apart from the trivial realisation $\widetilde{B}_a = \cO_X$ and $\widetilde{C}_a = 0$, one can also consider
\begin{equation}
\widetilde{B}_a = \cO_X(0,0,0,p_a)\oplus \cO_X(0,0,0,q_a) \ \ \ \text{and} \ \ \ \widetilde{C}_a = \cO_X(0,0,0,p_a+q_a)
\end{equation}
where $p_a$ and $q_a$ are positive integers. For this choice, the map $f_a = (f_{1,a},f_{2,a})$ contains two polynomials of multi-degrees $(0,0,0,p_a)$ and $(0,0,0,q_a)$. For generic choices of the polynomials, this map has rank one generically. The rank reduces to zero at points in $\IC\IP^1$ where $f_{1,a}=f_{2,a}=0$ but for sufficiently generic polynomials these equations have no solution in $\IC\IP^1$. Hence, this indeed provides a monad representation of the structure sheaf. Of course, the integers $p_a$, $q_a$ in $\widetilde{B}_a$ and $\widetilde{C}_a$ can appear in any of the four entries, so that we have a large number of choices on how to represent the structure sheaf as a monad. We can choose the trivial representation or a non-trivial representation characterized by choosing one of the four line bundle components and two integers $p_a$, $q_a$.

\vspace{12pt}
Now consider a given line bundle sum $V=\bigoplus_{a=1}^n L_a$. We can obtain monad representations for the individual line bundles by simply twisting the monad sequence~\eqref{eq:structure_sheaf} with $L_a$. This leads to
\begin{equation}
0\ \longrightarrow\ L_a \longrightarrow\ L_a\otimes \widetilde{B}_a\ \stackrel{f_a}{\longrightarrow}\ L_a\otimes \widetilde{C}_a\ \longrightarrow\ 0\; .  
\end{equation}
For the full line bundle sum $V$, we sum these sequences to obtain
\begin{equation}
0\ \longrightarrow\ V \longrightarrow\ B\ \stackrel{f}{\longrightarrow}\ C\ \longrightarrow\ 0  
\vspace{-21pt}
\end{equation}
where 
\begin{equation} \label{BtCt}
B = \bigoplus_{a=1}^n L_a\otimes \widetilde{B}_a\; ,\quad  C = \bigoplus_{a=1}^n L_a\otimes \widetilde{C}_a \; ,\quad f=\text{diag}\left(f_1, \ldots f_n \right)\; .
\end{equation}
We note that for each line bundle, $a$, we can choose $\widetilde{B}_a$ and $\widetilde{C}_a$ independently, from the range of possibilities explained above, so there is significant flexibility in the construction. For the diagonal form of the monad map $f$, as given above, each such choice leads to a monad representation of the original line bundle sum $V$. However, the map $f$ may allow deformations away from this block-diagonal form and then defines a more general class of bundles
\begin{equation}
0\ \longrightarrow\ \widetilde{V} \longrightarrow\ B\ \stackrel{f}{\longrightarrow}\ C\ \longrightarrow\ 0  \label{m1}
\end{equation}
which split into $V$ at the locus where $f$ becomes block-diagonal. Since ${\rm ch}(\tilde{V})={\rm ch}(B)-{\rm ch}(C)={\rm ch}(V)$ the Chern character of the monad bundle $\tilde{V}$ is the same as that of the original line bundle sum $V$. Therefore, if $c_1(V)$ vanishes and the line bundle sum $V$ satisfies the anomaly constraint~\eqref{anom} then the same is true for the monad bundle $\tilde{V}$.  
}

{\setstretch{1.15}
\subsubsection{Application to our example}
We would now like to apply the above procedure to our example on the tetra-quadric which was defined by a line bundle sum $V=\bigoplus_{a=1}^5L_a$ characterized by the integers
\begin{equation}\label{eq:example1}
(k_a^i)=
\cicy{ \\ \\ \\ \\ }
{ -1 & -1 & ~~0 & ~~1 & ~~1~ \\
~~0 & -3 & ~~1 & ~~1 & ~~1~ \\
~~0 & ~~2 & -1 & -1 & ~~0 ~\\
 ~~1 & ~~2 &~~ 0 & -1 & -2 ~\\}\; ,
\end{equation}
where the columns correspond to the line bundles $L_a$, $a=1,\ldots ,5$. To do this, we have to choose, for each $a=1,\ldots ,5$, the line bundle sums $\widetilde{B}_a$ and $\widetilde{C}_a$ which appear in the monad representation~\eqref{eq:structure_sheaf} of the structure sheaf. Our choice is
\begin{equation}
\begin{array}{lllllll}
\widetilde{B}_1&=&{\cal O}_X&,\quad&\widetilde{C}_1&=&0\\
\widetilde{B}_2&=&{\cal O}_X(0,2,0,0)^{\oplus 2}&,\quad&\widetilde{C}_2&=&{\cal O}_X(0,4,0,0)\\
\widetilde{B}_3&=&{\cal O}_X&,\quad&\widetilde{C}_3&=&0\\
\widetilde{B}_4&=&{\cal O}_X&,\quad&\widetilde{C}_4&=&0\\
\widetilde{B}_5&=&{\cal O}_X(0,0,0,2)^{\oplus 2}&,\quad&\widetilde{C}_2&=&{\cal O}_X(0,0,0,4)\\
\end{array}
\end{equation} 
From Eq.~\eqref{BtCt} this leads to line bundle sums $B=\bigoplus_{\alpha=1}^7{\cal O}_X({\bf b}_\alpha)$ and $C=\bigoplus_{\mu=1}^2{\cal O}_X({\bf c}_\mu)$ in the monad sequence~\eqref{m1} characterized by the integers
\beq\label{eq:bandc}
(b_\alpha^i)~=~~
\cicy{ \\ \\ \\ \\ }
{ - 1 & -1 & -1 & ~~0 &~~1& ~~1& ~~1~ \\
 ~~0 & -1 & -1 & ~~1 &~~1& ~~1& ~~1~ \\
~~0 & ~~2 & ~~2 & -1 &-1 &~~0 &~~0~\\
 ~~1 & ~~2 &~~ 2 & ~~0 &-1& ~~0 &~~0~\\}\
\hskip0.35in
(c_\mu^i)~=~~
\cicy{ \\ \\ \\ \\ }
{ - 1 & ~~1~ \\
 ~~1 & ~~1~ \\
~~2 & ~~0 ~\\
 ~~2 & ~~2~\\}\; .
\eeq
The general structure of the monad map is
\begin{equation}\label{monadmap}
 f\sim\left(\begin{array}{lllllll}f_{(0,1,2,1)}&f_{(0,2,0,0)}&f_{(0,2,0,0)}'&0&0&0&0\\
 f_{(2,1,0,1)}&0&0&f_{(1,0,1,2)}&f_{(0,0,1,3)}&f_{(0,0,0,2)}&f_{(0,0,0,2)}'\end{array}\right)\; ,
\end{equation} 
where the subscripts indicate the multi-degrees of the polynomials. For
\begin{equation}\label{splitcond}
 f_{(0,1,2,1)}=f_{(2,1,0,1)}=f_{(1,0,1,2)}=f_{(0,0,1,3)}=0
\end{equation}
the map is block-diagonal and $\widetilde{V}$ splits into the original line bundle sums $V$ so the coefficients in those polynomials parametrize the deformations away from the split locus.

It is important to point out that, even though most of our discussion will be carried out on the cover manifold, the line bundle sums $B$, $C$ in Eq.~\eqref{eq:bandc} are equivariant under the $\IZ_2\times\IZ_2$ symmetry of the tetra-quadric which we have used for our line bundle model and which has been defined in \eqref{Z2Z2def}. This means that, subject to an appropriate restriction of the monad map $f$, the monad bundle $\widetilde{V}$ has a $\IZ_2\times\IZ_2$ equivariant structure and descends to the quotient manifold.

Before we discuss any further properties, it is crucial to check whether the so-defined bundle $\widetilde{V}$ is superysmmetric, that is, whether it is poly-stable with slope zero. This is certainly the case at the locus where $\widetilde{V}$ splits into the line bundle sum $V$ (provided the K\"ahler moduli are chosen along the diagonal $t_1=t_2=t_3=t_4$ where all line bundle slopes vanish) but away from the split locus this is no longer obvious and has to be checked. 

\subsubsection{Stability of Monad Bundles}
Before we consider the monad bundle $\widetilde{V}$, we would like to review the general definition of poly-stability and outline the algorithm for checking it. 

First, we define the slope of a coherent sheaf $F$ by 
\begin{equation}
 \mu_{\bf t}(F) = \frac{1}{\text{rk}(F)} \int_X c_1(F)\wedge J\wedge J  = d_{ijk}\, c_1^i(F)\, t^j\, t^k = c_1^i(F)\,\kappa_i
\end{equation}
Note that this definition depends on the K\"ahler class $J=t^iJ_i$. A bundle $\widetilde{V}$ is said to be stable for a K\"ahler class ${\bf t}$ if $\mu_{\bf t}(F)<\mu_{\bf t}(\widetilde{V})$ for all sub-sheafs $F\subset \widetilde{V}$ with $0 < \text{rk}(F) < \text{rk}(\widetilde{V})$. Note that, due to the rank restriction on $F$, a line bundle is automatically stable. Further, a bundle $\widetilde{V}$ is poly-stable if it is a direct sum of stable bundles, $\widetilde{V}=U_1\oplus U_2\oplus\dots\oplus U_n$, all with the same slope, $\mu_{\bf t}(U_1)=\mu_{\bf t}(U_2)=\dots =\mu_{\bf t}(U_n)=\mu_{\bf t}(\widetilde{V})$. The bundle $\widetilde{V}$ is supersymmetric if it is poly-stable and has slope zero. 

Frequently, and indeed for our present example, we are dealing with bundles $\widetilde{V}$ with vanishing first Chern class, so that $\mu_{\bf t}(\widetilde{V})=0$ automatically. In this case, poly-stability requires that $\mu_{\bf t}(U_r)=0$, for all $r=1,\ldots ,n$,   and that $\mu_{\bf t}(F)<0$ for all sub-sheafs $F\subset U_r$ with $0<{\rm rk}(F)<{\rm rk}(U_r)$.

For a poly-stable bundle $\widetilde{V}$, its dual $\widetilde{V}^{^*}$ and $\widetilde{V} \otimes L$ for any line bundle $L$ are also poly-stable. A stable bundle $\widetilde{V}$ with $\text{rk}(\widetilde{V})>1$ and vanishing slope must satisfy $H^0(X, \widetilde{V}) = H^3(X,\widetilde{V}) = 0$ as well as $H^0(X,\wedge^k V) = H^3(X,\wedge^k V) = 0$.
While the slope of a bundle can be easily computed, finding all coherent sub-sheafs of a given bundle is difficult. A practical algorithm which deals with this complication is as follows. For a sub-sheaf $F$ of $\widetilde{V}$, define $k={\rm rk}(F)$ and the line bundle $L=\wedge^kF$. Then $L$ is a sub-sheaf of $\wedge^k\widetilde{V}$ and $\mu_{\bf t}(L)=k\mu_{\bf t}(F)$. So the region in the K\"ahler cone $C_{\bf t}$ where the bundle $\widetilde{V}$ (with $c_1(\widetilde{V})=0$) is supersymmetric is
\begin{equation}
 C_{\widetilde{V}}=\{{\bf t}\in C_{\rm t}\,|\, \mu_{\bf t}(L)\leq 0\mbox{ for all line bundles }L\mbox{ which inject into }\wedge^k\widetilde{V}\; ,k=1,\ldots,{\rm rk}(\widetilde{V})-1\}\; . \label{CV}
\end{equation}  
Note that, while stability requires $\mu_{\bf t}(L)<0$, we have used $\mu_{\bf t}(L)\leq 0$ in the above definition. In fact, at a locus in K\"ahler moduli space where $\mu_{\bf t}(L)=0$ the bundle splits into a direct sum and is no longer stable but poly-stable and, hence, still supersymmetric. In Ref.~\cite{Anderson:2009sw} this has been referred to as a stability wall. In the present context, this is of course precisely what we expect to happen when the bundle $\widetilde{V}$ splits into a line bundle sum $V$. 
The detailed procedure to find the supersymmetric region based on Eq.~\eqref{CV} has been developed in \cite{Anderson:2008uw, Anderson:2008ex} and summarised in Appendix~\ref{sec:stab_criteria}.

\subsubsection{Checking stability for our example}\label{sec:stabex}
In order to compute the supersymmetric region~\eqref{CV} for our bundle $\widetilde{V}$ on the tetra-quadric, it is useful to describe the K\"ahler cone in terms of the variables $s_i=\kappa_i=d_{ijk}\,t^j\,t^k$. We have seen in Eq.~\eqref{kcs} that, in those variables, the K\"ahler cone is
\beq
 C_{\bf s}=\{\mathbf{s}\in \mathbb R ^4 \ | \ \mathbf{s.e}_i\geq 0 \text{ and } \mathbf{s.n}_i\geq 0 \}\; .
 \eeq
Then, for a line bundle $L={\cal O}_X(-{\bf k})$, the slope is simply given by the dot product $\mu_{\bf s}(L)=-{\bf s}\cdot{\bf k}$. The supersymmetric region, expressed in terms of the ${\bf s}$ coordinates, can, therefore be written as
\begin{equation}
C_{\widetilde{V}}=\{\mathbf{s}\in C_{\bf s}\, |\, \mathbf{s}.\mathbf{k} \geq 0 \text{ for any } L = \mathcal{O}(-\mathbf{k}),\, L \text{ injects into } \wedge^k \widetilde{V},\ k = 1, \ldots , \text{rk}(\widetilde{V} )-1 \} \; . 
\end{equation}
Note that this is an intersection of hyperplanes so the supersymmetric region (written in the ${\bf s}$ coordinates) forms a cone. 

We now need to find all line bundles, $L$, on the tetra-quadric which inject into some wedge power of the monad bundle $\widetilde{V}$ defined by Eqs.~\eqref{eq:bandc}. Appendix~\ref{sec:stab_criteria} sets out a number of simple sufficient conditions, based on computing cohomology dimensions only, for line bundles to inject or not to inject. Specifically, these criteria are given in (\ref{eq:cohcondiiton1}), (\ref{eq:cohcondiiton2}), (\ref{eq:cohcondiiton3}), (\ref{eq:cohcondiiton4}), (\ref{eq:cohcondiiton5}). Carrying out a scan over all line bundles with entries in the range from $-3$ to $3$ we collect all line bundles which definitely inject into some power of $\widetilde{V}$, according to our criteria. These line bundles reduce the supersymmetric region to the following: 
\begin{equation}\label{eq:hyperplane}
  C_{\widetilde{V}}=\{\mathbf{s}\in C_{\bf s}\, |\, \ \mathbf{s.}(1,1,-1,-1)= 0 \text{ and } \mathbf{s.}(-1,1,0,0)\geq 0  \text{ and } \mathbf{s.}(1,1,-2,0)\geq 0 \}\; .
\end{equation} 
 Increasing the range of line bundles integers further did not change this region of stability. For some line bundles the simple numerical criteria in Appendix~\ref{sec:stab_criteria} are not sufficient to decide whether or not they inject. In such cases, we have to compute ranks of relevant maps which is more involved. We will return to this problem in Section~\ref{sec:stability2} but we note here that, as it turns out, these line bundles do not change the result~\eqref{eq:hyperplane}.

The first condition in Eq.~\eqref{eq:hyperplane} comes from the fact that the line bundle $L_4={\cal O}_X(1,1,-1,-1)$ injects into $\widetilde{V}$, while its dual $L_4^{^*}$ injects into $\wedge^4 V\cong V^{^*}$. This means that we are confined to the hyperplace $s_1+s_2=s_3+s_4$ in K\"ahler moduli space and that we have a poly-stable split bundle $\widetilde{V}=U\oplus L_4$ which corresponds to the locus in bundle moduli space where the entry $f_{(0,0,1,3)}$ in the monad map~\eqref{monadmap} vanishes. The locus in bundle moduli space where equality holds for all three conditions in \eqref{eq:hyperplane}, which corresponds to $t_1=t_2=t_3=t_4$, is precisely where the conditions~\eqref{splitcond} have to be satisfied and the bundle fully splits into the line bundle sum~\eqref{lbs}. Note that $L_4$ is indeed one of the line bundles defining the original line bundle sum~\eqref{lbs}. Therefore, in its supersymmetric moduli space, $\widetilde{V}$ does not provide a fully non-Abelian version of this line bundle sum but only ``connects" four of the five line bundles. In terms of the effective field theory language, the locus in bundle moduli space where $L_4$ remains split off corresponds to vanishing VEVs for all ${\bf 1}_{a,b}$ which carry a $4$--index. From the spectrum~\eqref{gutspec}, these are precisely the singlets ${\bf 1}_{2,4}$. Our earlier arguments based on the effective field theory suggested that the three vector-like ${\bf 5}-\overline{\bf 5}$ multiplets (which give rise to the Higgs doublets) remain massless whenever $\langle {\bf 1}_{2,4}\rangle =0$ and we have verified this expectation explicitly by computing the cohomology of the extension bundles. Here we see that $\langle {\bf 1}_{2,4}\rangle =0$ in the entire supersymmetric moduli space of our monad bundle $\widetilde{V}$ so we expect three massless ${\bf 5}-\overline{\bf 5}$ pairs from the cohomology of~$\wedge^2\widetilde{V}$. We will now verify this expectation by an explicit compution.

\subsubsection{The Spectrum}\label{sec:monadspectrum}
In this subsection we will discuss the spectrum of the $S(U(4)\times U(1))$ compactification with the monad bundle $\widetilde{V}=U\oplus L_4$. The low-energy theory has a gauge group $SU(5)$ and an additional $U(1)$ symmetry, which is massive and, hence, global. As discussed in the next subsection, this additional symmetry, henceforth referred to as $U_X(1)$, combined with the hypercharge leads to the well-known $B-L$ symmetry.

\vspace{12pt}
We will now compute the cohomology of the bundle $\widetilde{V}$. We begin by writing down the long exact sequence

\begin{equation}
 \begin{array}{cccccc}
 &\widetilde{V}&\longrightarrow& B&\longrightarrow& C\\[8pt]
 h^0(X,\cdot\,)$\ \ \ \ $&h^0(X,\widetilde{V})&&8&&12\\[3pt]
 h^1(X,\cdot\,)$\ \ \ \ $&h^1(X,\widetilde{V})&&8&&0\\[3pt]
 h^2(X,\cdot\,)$\ \ \ \ $&h^2(X,\widetilde{V})&&0&&0\\[3pt]
 h^3(X,\cdot\,)$\ \ \ \ $&h^3(X,\widetilde{V})&&0&&0
\end{array}\; 
\label{e1.1}
\vspace{12pt}
\end{equation} 
associated to the monad sequence~\eqref{eq:monad}. Using an explicit representation for the cohomology groups $H^0(X,B)$ and $H^0(X,C)$, as well as for the map between these groups induced by the monad map, it follows that~\footnote{These results also follow from $h^0(X,\widetilde{V})=0$ which is a consequence of the poly-stability of $\widetilde{V}$.}

\begin{equation}
\begin{aligned}
\text{dim}\, \text{Ker} & \left( H^0(X,B)  \longrightarrow H^0(X,C)  \right) = 0\\
\text{dim} \, \text{Coker} & \left( H^0(X,B)  \longrightarrow H^0(X,C)  \right) = 4\; .
 \end{aligned}
 \label{e1.2}
 \vspace{12pt}
\end{equation}
This implies
\begin{equation}
 h^{^{\!\bullet}}(X,V)=(0,12,0,0)\;,
 \label{e1.3}
\end{equation} 
and, hence, a spectrum with $12$ ${\bf 10}$ multiplets and no $\overline{\bf 10}$ mutliplets, the same as we found for the original line bundle sum $V$ in Eq.~\eqref{cohV}. Of course, this does not come as a surprise, since these $12$ chiral multiplets are protected by the index. Also note that $h^1 (X, \widetilde{V})= h^1(X, U)+ h^1 (X, L_4)$. Since $h^1 (X, L_4)=0$ \footnote{In fact the line bundle $L_4= {\cal O}_X (1, 1, -1, -1)$ has entirely vanishing 
cohomologies.} we get $h^1 (X, \widetilde{V})= h^1(X, U)=12$, so the multiplets transforming as ${\bf 10}$ come from the cohomology group $H^1(X, U)$.

\vspace{21pt}
It is considerably more interesting -- and difficult -- to analyze the spectrum of ${\bf 5}-\overline{\bf 5}$ multiplets which follows from $\wedge^2\widetilde{V}$. To do this, we explicitly focus on the supersymmetric moduli space where the bundle splits as $\widetilde{V}=U\oplus L_4$, and $U$ is described by the monad
\begin{equation}\label{eq:monadseqnew}
0\ \longrightarrow\ U \longrightarrow\ \widetilde{B}\ \stackrel{\tilde{f}}{\longrightarrow}\ C\ \longrightarrow\ 0 \; ,
\end{equation}
with the line bundle sums $\widetilde{B}$ and $C$ defined by the bundle integers
\vspace{12pt}
\beq\label{eq:bandc2}
(\tilde{b}_\alpha^i)~=~~
\cicy{ \\ \\ \\ \\ }
{ - 1 & -1 & -1 & ~~0 & ~~1& ~~1~ \\
 ~~0 & -1 & -1 & ~~1& ~~1& ~~1~ \\
~~0 & ~~2 & ~~2 & -1 &~~0 &~~0~\\
 ~~1 & ~~2 &~~ 2 & ~~0 & ~~0 &~~0~\\}\
\hskip0.35in
(c_\mu^i)~=~~
\cicy{ \\ \\ \\ \\ }
{ - 1 & ~~1~ \\
 ~~1 & ~~1~ \\
~~2 & ~~0 ~\\
 ~~2 & ~~2~\\}\; .
\vspace{14pt}
\eeq
Note these are the same line bundle sums as in the original monad~\eqref{eq:bandc} except that we have removed the fifth column from $(b_\alpha^i)$ which corresponds to the line bundle $L_4={\cal O}_X(1,1,-1,-1)$. The corresponding monad map $\tilde{f}$ is obtained by likewise removing the fifth column from the original monad map $f$ in Eq.~\eqref{monadmap} which results in

\begin{equation}\label{monadmap1}
 f\sim\left(\begin{array}{llllll}f_{(0,1,2,1)}&f_{(0,2,0,0)}&f_{(0,2,0,0)}'&0&0&0\\
 f_{(2,1,0,1)}&0&0&f_{(1,0,1,2)}&f_{(0,0,0,2)}&f_{(0,0,0,2)}'\end{array}\right)\; ,
\vspace{12pt}
\end{equation} 
The locus where $U$ splits into the line bundle sum $L_1\oplus L_2\oplus L_3\oplus L_5$ is $t_1=t_2=t_3=t_4$ in K\"ahler moduli space and 
\begin{equation}
 f_{(0,1,2,1)}=f_{(2,1,0,1)}=f_{(1,0,1,2)}=0
\end{equation}
in bundle moduli space.  

\vspace{31pt}
We need to compute the cohomology of $\wedge^2\widetilde{V} = \left(L_4\otimes U\right) \, \oplus\, \wedge^2 U$. The cohomology of $L_4\otimes U$ can be easily obtained by twisting the monad sequence (\ref{eq:monadseqnew}) with $L_4$:
\begin{equation}
0\ \longrightarrow\ L_4\otimes U \longrightarrow\ L_4\otimes \widetilde{B}\ \longrightarrow\ L_4\otimes C\ \longrightarrow\ 0  \,. 
\label{e1.4}
\end{equation}
From the associated long exact sequence, together with $h^{^{\!\bullet}}(X,L_4\otimes\widetilde{B})=(8,8,0,0)$ and $h^{^{\!\bullet}}(X,L_4\otimes C)=(12,0,0,0)$, it follows that

\beq
 h^{^{\!\bullet}}(X,L_4\otimes U)\ \,  =\  (0,12,0,0)  
 \label{e1.5}
\vspace{12pt}
\eeq
so this part of $\wedge^2\widetilde{V}$ takes care of the chiral asymmetry. The fate of the Higgs doublets is determined entirely by $\wedge^2U$ and in order to compute this part we need to use the second exterior power of the monad sequence~(\ref{eq:monadseqnew}):
\vspace{10pt}
\beq
0\ \longrightarrow\ \wedge^2U \longrightarrow\ \wedge^2\widetilde{B}\ \longrightarrow\ \widetilde{B}\otimes C\ \longrightarrow\ S^2C\ \longrightarrow\ 0
\vspace{31pt}
\eeq
As usual, this can be split in two short exact sequences, whose associated long exact sequences read:
\vspace{10pt}
\begin{equation}\label{eq:2sequences}
  \begin{array}{cccccccccccccc}
 &\wedge^2U&\longrightarrow& \wedge^2\widetilde{B}&\longrightarrow& Q&\quad\quad\quad\quad&Q&\longrightarrow&\widetilde{B}\otimes C&\longrightarrow&S^2C\\[8pt]
 h^0(X,\cdot\,)$\ \ \ \ $&h^0(\wedge^2U)&&53&&h^0(Q)&& h^0(Q)&&150&&96\\[3pt]
 h^1(X,\cdot\,)$\ \ \ \ $& h^1(\wedge^2U) &&85&&h^1(Q)&&h^1(Q)&&134&&48\\[3pt]
 h^2(X,\cdot\,)$\ \ \ \ $&h^2(\wedge^2U)&&0&&h^2(Q)&& h^2(Q)&&0&&0\\[3pt]
 h^3(X,\cdot\,)$\ \ \ \ $&h^3(\wedge^2U)&&0&&h^3(Q)&& h^3(Q)&&0&&0
\end{array}
\end{equation} 

The second short exact sequence implies that 
\begin{equation}
\begin{aligned}
H^0(X,Q) & \cong \text{Ker}\left( H^0(X,\widetilde{B}\otimes C) \longrightarrow H^0(X,S^2C)  \right)\\ 
H^1(X,Q) & \cong \text{Coker}  \left( H^0(X,\widetilde{B}\otimes C) \longrightarrow H^0(X,S^2C)  \right) \\
 & \oplus \text{Ker}\left( H^1(X,\widetilde{B}\otimes C) \longrightarrow H^1(X,S^2C)  \right) \\ 
H^2(X,Q) & \cong \text{Coker}\left( H^1(X,\widetilde{B}\otimes C) \longrightarrow H^1(X,S^2C)  \right) \\
H^3(X,Q) & \cong 0
\end{aligned}
\eeq
The computation of these cohomology groups proceeds in several stages. In the first step, we need to find the map between the line bundle sums $\widetilde{B}\otimes C$ and $S^2C$ induced by the monad map (\ref{eq:monadseqnew}). In the second step, we construct the induced map between the various cohomology groups and compute their ranks using the CICY package \cite{CICYpackage}. A comprehensive exposition on the computation of cohomology groups and of ranks of maps goes beyond the scope of the present paper. The interested reader can find in Ref.~\cite{Anderson:2013qca} an outline of the basic techniques for computing line bundle cohomology on complete intersection Calabi-Yau manifolds in products of projective spaces. 

\vspace{12pt}
We find that the map $H^0(X,\widetilde{B}\otimes C) \longrightarrow H^0(X,S^2C)$ has rank $94$,  while the map between $H^1(X,\widetilde{B}\otimes C) \longrightarrow H^1(X,S^2C)$ has maximal rank $48$. This leads to 
\beq
 h^{^{\!\bullet}}(X,Q)\ \,  =\  (56,88,0,0)  
\eeq

The final step consists in determining the cohomology of $\wedge^2  U$. The first long exact sequence in cohomology in (\ref{eq:2sequences}) implies:
\begin{equation}
\begin{aligned}
H^0(X,\wedge^2 U) & \cong \text{Ker}\left( H^0(X,\wedge^2\widetilde{B}) \longrightarrow H^0(X,Q)  \right)\\ 
H^1(X,\wedge^2 U) & \cong \text{Coker}  \left( H^0(X,\wedge^2\widetilde{B}) \longrightarrow H^0(X,Q)  \right) \\
 & \oplus \text{Ker}\left( H^1(X,\wedge^2\widetilde{B}) \longrightarrow H^1(X,Q)  \right) \\ 
H^2(X,\wedge^2 U) & \cong \text{Coker}\left( H^1(X,\wedge^2\widetilde{B}) \longrightarrow H^1(X,Q)  \right) \\
H^3(X,\wedge^2U) & \cong 0
\end{aligned}
\eeq
Computing these maps involves several layers of additional complication. To start with, the map $H^0(X,\wedge^2\widetilde{B}) \longrightarrow H^0(X,Q)$ is induced by the bundle map $\wedge^2\widetilde{B} \longrightarrow Q$ which itself has to be determined from the monad map (\ref{eq:monadseqnew}). However, since 
\beq
H^0(X,Q)  \cong \text{Ker}\left( H^0(X,\widetilde{B}\otimes C) \longrightarrow H^0(X,S^2C)  \right)
\eeq
it follows that $H^0(X,Q)$ is a subspace of $H^0(X,\widetilde{B}\otimes C)$ and thus the map $H^0(X,\wedge^2\widetilde{B}) \longrightarrow H^0(X,Q)$ is equivalent with the map $H^0(X,\wedge^2\widetilde{B}) \longrightarrow H^0(X,\widetilde{B}\otimes C)$, with the single difference that, for the latter, the target space is larger. Computing the rank of $H^0(X,\wedge^2\widetilde{B}) \longrightarrow H^0(X,Q)$, we obtain 53. This leaves us with the following tableaux of dimensions

\beq
\begin{array}{ccccccccc}
0 \ \ &   \longrightarrow & \wedge^2 U &  \longrightarrow  &\wedge^2 \widetilde{B} & \longrightarrow& Q & \longrightarrow &0\ \ \ \ \\   \\[-12pt]
& & 0 &&53&& 56 &&  \\
& & 3+K &&85&& 88 & & \\
& & C &&0&&0&&  \\
& & 0 &&0&&0&&  \\
\end{array}
\eeq
where $K = \text{dim}\, \text{Ker}\left( H^1(X,\wedge^2\widetilde{B}) \longrightarrow H^1(X,Q)  \right)$ and $C = \text{dim}\, \text{Coker}\left( H^1(X,\wedge^2\widetilde{B}) \longrightarrow H^1(X,Q)  \right)$. From exactness, it follows that $C = 3 + K$. Computing the map $H^1(X,\wedge^2\widetilde{B}) \longrightarrow H^1(X,Q)$ would require the knowledge of a co-boundary map and we would have to deal with the additional complication that the target space $H^1(X,Q)$ is a direct sum of $H^0(X,S^2C)$ and $H^1(X, \widetilde{B}\otimes C)$. Fortunately, for the present case, the information acquired so far is enough to make an important statement. We have obtained that
\beq
 h^{^{\!\bullet}}(X,\wedge^2 \widetilde{V}) \ \, = \  h^{^{\!\bullet}}(X, L_4\otimes U) \ +\  h^{^{\!\bullet}}(X,\wedge^2 U) \ \,  =\  (0,15+K,3+K,0)  
\eeq
where $K\geq 0$. We recall that the corresponding cohomology at the split locus (where $\widetilde{V}$ splits into the line bundle sum $V$ in \eqref{lbs}) is given by $h^{^{\!\bullet}}(X,V)=(0,15,3,0)$. This means that the three massless, vector-like ${\bf 5}-\overline{\bf 5}$ states present at the split locus remain massless throughout the supersymmetric moduli space of $\widetilde{V}$, as expected from our low-energy arguments. 
In principle, the above cohomology calculation still allows $K>0$. However, from the viewpoint of the effective field theory, this is not expected. Indeed, as the move away from the split locus by switching on singlet VEVs we may generate mass terms for vector-like pairs, thereby reducing their number, but we do not expect this number to increase. Thus we conclude that $K=0$. 

\vspace{12pt}
In summary, our result guarantees the presence of massless ${\bf 5}-\overline{\bf 5}$ multiplets (resulting in Higgs doublets in the downstairs theory) for non-Abelian deformations of the original line bundle sum $V$ which are of the form $\widetilde{V}=L_4\oplus U$. This result is in perfect agreement with the expectation from the four-dimensional effective theory and the earlier cohomology computation in the context of extension bundles.

Finally, the spectrum also contains singlets (vector bundle moduli), which correspond to elements of the cohomology group 
\begin{equation}
H^1(X,\text{End}(\widetilde V)) = H^1(X, \widetilde V \otimes \widetilde V^{^*}) = H^1(X, U \otimes U^{^*}) \oplus H^1(X, L_4 \otimes U^{^*}) \oplus H^1(X, U \otimes L_4^{^*}) 
\end{equation}

The cohomology of $U\otimes L_4^{^*}$ can be easily obtained by twisting the monad sequence (\ref{eq:monadseqnew}) with $L_4^{^*}$. From the associated long exact sequence, together with the information $h^{^{\!\bullet}}(X,\widetilde{B}\otimes L_4^{^*})=(8,40,0,0)$ and $h^{^{\!\bullet}}(X, C\otimes L_4^{^*})=(8,28,0,0)$, it follows, after computing the necessary ranks of cohomology maps, that
\beq
 h^{^{\!\bullet}}(X, U\otimes L_4^{^*})\ \,  =\  (0,12,0,0)  
\eeq
Thus the number of singlets is given by:
\beq
 h^1(X, \widetilde V\otimes \widetilde V^{^*})\ \,  =\  12\, +\,  h^1(X, U\otimes U^{^*})\label{VUrel}
\eeq

The cohomology of $U\otimes U^{^*}$ can be computed using a web of six short exact sequences. Unfortunately, this computation runs into the same difficulties as encountered in the computation of $\mathbf{5}-\overline{\mathbf{5}}$ pairs discussed above and we will not carry it out explicitly. However, there is an expectation for this cohomology from low-energy arguments. As we move away from the Abelian locus, three of the four $U(1)$ symmetries are broken spontaneously and, as a result, three bundle moduli should acquire a mass. With $60$ bundle moduli at the Abelian locus we, therefore, expect that $h^1(X, \widetilde V\otimes \widetilde V^{^*})=57$ and, from Eq.~\eqref{VUrel}, that  $h^1(X, U\otimes U^{^*})=45$. 

As explained in the next section, the 12 moduli coming from $ h^1(X, U\otimes L_4^{^*})$ are charged under the extra $U(1)$ symmetry. As such, a non-trivial vacuum value for these singlets breaks the $U_X(1)$ and corresponds to moving away from the $S(U(4) \times U_X(1))$ locus into the moduli space of generic $SU(5)$ bundles. These are the moduli ${\bf 1}_{2,4}$ in the spectrum~\eqref{gutspec}. The remaining singlets from  $ H^1(X, U\otimes U^{^*})$ are uncharged under $U_X(1)$. 


\subsubsection{The $U_X(1)$ symmetry and $B-L$}
In the present section we will show that the additional global $U_X(1)$ symmetry, combined with $U_Y(1)$ hypercharge, leads to the well-known $B-L$ symmetry. For this, we need to compute the $U_X(1)$ charges of the various multiplets in the low-energy GUT. We do this by considering the sequence of embeddings $S\left(U(4)\times U_X(1) \right) \subset SU(5)\subset E_8$. The matter multiplets can be obtained by decomposing the adjoint $\mathbf{248}_{E_8}$ of $E_8$ under the $SU(5)\times U_X(1)$ sub-group. The corresponding branching rule has been discussed in Ref.~\cite{Anderson:2012yf}, which we review below. 

The $U_X(1)$ charges of the GUT multiplets can be represented by vectors $\mathbf{q}$. Due to the determinant condition, two such vectors $\mathbf{q}$ and $\mathbf{\widetilde q}$ have to be identified if $\mathbf{q}-\widetilde{\mathbf{q}} = \IZ \mathbf{n}$, where $\mathbf{n}=(4,1)$. We summarise in Table~\ref{spectrum2} the resulting spectrum. The actual $U_X(1)$ charges can be recovered from this description by multiplying a charge vector $\mathbf{q}=(q_1,q_2)$ with $(-1,4)$.
\vspace{12pt}

\begin{table}[h]
\hspace{.25cm}
\parbox{.45\linewidth}{
\centering
\begin{tabular}{| c | c | r |}
\hline
\varstr{14pt}{9pt} repr. & cohomology & $U_X(1)$ charge  \\ \hline\hline
\varstr{14pt}{9pt} $~{\bf 1}_{0}$ & $H^1(X, U \otimes U^{^*})$  &  0~~~~~~~  \\ \hline
\varstr{14pt}{9pt} $~{\bf 1}_{\mathbf{e}_1-\mathbf{e}_2}$ & $H^1(X, U \otimes L^{^*})$  &  -5~~~~~~~  \\ \hline
\varstr{14pt}{9pt} $~{\bf 1}_{-\mathbf{e}_1+\mathbf{e}_2}$ & $H^1(X, L \otimes U^{^*})$  &  5~~~~~~~  \\ \hline

\varstr{14pt}{9pt} $~{\bf 5}_{-2\mathbf{e}_1}$ & $H^1(X, \wedge^2 U^{^*})$  & 2 ~~~~~~  \\ \hline
\varstr{14pt}{9pt} $~{\bf 5}_{-\mathbf{e}_1-\mathbf{e}_2}$ & $H^1(X,  U^{^*}\otimes L^{^*})$  & -3 ~~~~~~  \\ \hline
 \end{tabular}
}
\hspace{.5cm}
\parbox{.45\linewidth}{
\centering
\begin{tabular}{| c | c | r |}
\hline

\varstr{14pt}{9pt} $~~~{\bf \overline{5}}_{2\mathbf{e}_1}~~$ & $H^1(X, \wedge^2 U)$  & ~~~~~ -2~~~~~~  \\ \hline
\varstr{14pt}{9pt} $~{\bf \overline{5}}_{\mathbf{e}_1+\mathbf{e}_2}$ & $~H^1(X, U\otimes L)~$  & 3~~~~~~  \\ \hline
\varstr{14pt}{9pt} $~{\bf 10}_{\mathbf{e}_1}$ &$H^1(X, U)$ & -1~~~~~~  \\ \hline
\varstr{14pt}{9pt} $~{\bf 10}_{\mathbf{e}_2}$ &$H^1(X, L)$ & 4~~~~~~  \\ \hline
\varstr{14pt}{9pt} $~{\bf  \overline{10}}_{-\mathbf{e}_1}$ & $H^1(X, U^{^*})$ & 1~~~~~~ \\ \hline
\varstr{14pt}{9pt} $~{\bf  \overline{10}}_{-\mathbf{e}_2}$ & $H^1(X, L^{^*})$ &  -4~~~~~~ 
\\ \hline 

 \end{tabular}
}\vspace{10pt}
\begin{center}
\parbox{15cm}{\caption{\it\small The spectrum of $SU(5)$ GUT models derived from heterotic models with $S\left( U(4)\times U(1)\right)$ bundles $U\oplus L$. Here $\mathbf{e}_1$ and $\mathbf{e}_2$ are the standard unit vectors in two dimensions. The third column has been obtained by projecting the charges $\mathbf{q}$ along the line $(-1,4)$.}\label{spectrum2}}
\end{center}
\vspace{-21pt}
\end{table}

Looking at the dimensions of the various cohomology groups, as computed in the previous section, we obtain the following spectrum: 
\vspace{-4pt}
\begin{equation}
 {12\,\bf{10}}_{-1}\; ,\quad 12\,\overline{\bf 5}_{3}\;,\quad 3\,\overline{\bf 5}_{-2}\;,\quad 3\, {\bf 5}_{2}\; ,\quad 12\,{\bf 1}_{-5}\; ,\quad h^1(X, U\otimes U^{^*}) \,{\bf 1}_{0} 
\vspace{-4pt}
\end{equation} 
It is well-known, in the context of regular $SO(10)$ GUTs, that the $U(1)$ factor in the standard decomposition $SO(10)\rightarrow SU(5)\times U(1)$ leads to $B-L$ with $SO(10)$ multiplets branching as
\vspace{-4pt}
\begin{equation}
{\bf 16}_{SO(10)}= {\bf 1}_{-5} + {\bar {\bf 5}}_3 + {\bf 10}_{-1}\; ,\quad {\bf 10}_{SO(10)}= {\bf 5}_{2} + {\overline {\bf 5}}_{-2}\,. 
\label{e6}
\vspace{-4pt}
\end{equation}
The observation is that these $U(1)$ charges precisely coincide with the above $U_X(1)$ charges in our model. Explicitly, $B-L$ is given by the following combination
\vspace{-4pt}
\beq
B -L \ =\ -\frac{1}{5} X\, +\, \frac{2}{5} Y
\vspace{-4pt}
\eeq
of $U_X(1)$ and hypercharge. The absence of operators leading to fast proton decay in our $SU(5)\times U_X(1)$ model is in part, but not fully, explained by the presence of the $U_X(1)$ symmetry. For example, while dimension four operators of the form $\overline{\bf 5}\,\overline{\bf 5}\,{\bf 10}$ are clearly forbidden, dimension five operators $\overline{\bf 5}_3\,{\bf 10}_{-1}\,{\bf 10}_{-1}\,{\bf 10}_{-1}$ are allowed by the $U_X(1)$ symmetry. However, from Eq.~\eqref{Wgut}, these operators are forbidden, even at a general point in moduli space, by the additional $U(1)$ symmetries which arise at Abelian locus. 

Let us finish this section with the following comment. The $12$ singlets in our model, charged under $U_X(1)$, can be viewed as right-handed neutrinos in analogy with conventional $SO(10)$ GUTs. As it happens, their number equals the number of families so that the charged spectrum forms $12$ copies of ${\bf 16}_{SO(10)}$. This means that the $U_X(1)$ is non-anomalous. This should be considered a coincidence in this particular model, as, in general, the number of singlets does not have to equal the number of families. 

\vspace{-4pt}
\subsubsection{More on stability}\label{sec:stability2}
In Section~\ref{sec:stabex} we have argued that the bundle $\widetilde{V}$ is poly-stable in the region
\begin{equation}\label{eq:hyperplane1}
  C_{\widetilde{V}}=\{\mathbf{s}\in C_{\bf s}\, |\, \ \mathbf{s.}(1,1,-1,-1)= 0 \text{ and } \mathbf{s.}(-1,1,0,0)\geq 0  \text{ and } \mathbf{s.}(1,1,-2,0)\geq 0 \}\; .
\end{equation} 
in K\"ahler moduli space and that, due to the first condition above, it splits into a poly-stable bundle $\widetilde{V}=U\oplus L_4$. This result was based on computing cohomology dimensions only. However, for a complete stability analysis we need to consider line bundles injecting into all wedge powers $\wedge^k\widetilde{V}$, where $k=1,\ldots ,4$, by computing ranks of relevant maps. In this sub-section, we take up these computations and show that they do not change the above result for the poly-stable region. 

Given that our bundle is already split, this task is somewhat simplified. All we need to consider is the line bundles injecting into powers of the rank four bundle $U$, that is, into $U$, $\wedge^2 U$ and $\wedge^3U$, which is the case for a line bundle $L$, precisely when $H^0(X,\wedge^kU\otimes L^{^*})$ is non-trivial. Starting with $U$ itself, a line bundle $L$ injects into it precisely if 
\beq
 H^0\left(X, U \otimes L^{^*} \right) \cong \text{Ker}\left( H^0\big(\widetilde{B}\otimes L^{^{\!*}}\big)\ \longrightarrow\ H^0\big(C\otimes L^{^{\!*}}\big)\right)
\eeq
is non-trivial. We have computed this kernel for all the line bundles with entries between $-3$ and $3$ and have obtained the following set of injecting line bundles:
\begin{table}[!h]
\vspace{4pt}
\begin{center}
\begin{tabular}{ c c c c c }
\varstr{14pt}{9pt} (-3, -3, -3, 2), &(-3, -3, -2, 2), &(-3, -3, -1, 2), &(-3, -3, 0, 1), &(-3, -3, 0, 2) \\ 
\varstr{14pt}{9pt}(-3, -2, -3, 1), &(-3, -2, -2, 1), &(-3, -1, -3, 1), &(-3, 1, -3, -1), &(-3, 1, -2, -1), \\
\varstr{14pt}{9pt}(-3, 1, -1, -3), &(-3, 1, -1, -2), &(-3, 1, -1, -1), &(-3, 1, 0, -3), &(-3, 1, 0, -2), \\
\varstr{14pt}{9pt}(-3, 2, -3, -3), &(-2, -3, -3, 2), &(-2, -3, -2, 2), &(-2, -3, -1, 2), &(-2, -3, 0, 1), \\
\varstr{14pt}{9pt}(-2, -3, 0, 2), &(-2, -2, -3, 1), &(-2, -1, -3, 1), &(-2, 1, -3, -1), &(-2, 1, -2, -1), \\
\varstr{14pt}{9pt}(-2, 1, -1, -3), &(-2, 1, -1, -2), &(-2, 1, -1, -1), &(-2, 1, 0, -3), &(-2, 1, 0, -2), \\
\varstr{14pt}{9pt}(-1, -3, -3, 1), &(-1, -3, -3, 2), &(-1, -3, -2, 1), &(-1, -3, -2, 2), &(-1, -3, -1, 1), \\
\varstr{14pt}{9pt}(-1, -3, -1, 2), &(-1, -3, 0, 1), &(-1, -3, 0, 2), &(-1, 1, -3, -1), &(-1, 1, -2, -1), \\
\varstr{14pt}{9pt}(-1, 1, -1, -3), &(-1, 1, -1, -2), &(-1, 1, -1, -1), &(-1, 1, 0, -3), &(-1, 1, 0, -2)
\end{tabular}
 \end{center}
 \vspace{-12pt}
 \end{table}
\newline However, none of these line bundles de-stabilises the cone $C_{\widetilde{V}}$. To find the line bundles $L$ injecting into $\wedge^2 U$, we need to find the non-trivial cohomology groups 
\beq
 H^0\left(X, \wedge^2U \otimes L^{^*} \right) \cong \text{Ker}\left( H^0\big(\wedge^2\widetilde{B}\otimes L^{^{\!*}}\big)\! \longrightarrow \text{Ker}\big( H^0\big(\widetilde{B}\otimes C\otimes L^{^{\!*}}\big)\! \longrightarrow\! H^0\big(S^2C\otimes L^{^{\!*}}\big)\big)\!\right)\; .
\eeq
This computation is similar to that performed in Section~\ref{sec:monadspectrum} in order to decide the existence of $\mathbf{5}-\overline{\mathbf{5}}$ pairs. Although computationally challenging, we have computed the cohomology $H^0\left(X, \wedge^2 U \otimes L^{^*} \right)$ for all the line bundles with entries between $-1$ and $1$ and we find no injecting line bundles.  

\vspace{20pt}
Finally, for the line bundles that potentially inject into $\wedge^3 U$ we need to compute cohomology groups of the type $H^0\left( X, \wedge^3 U\otimes L^{^*} \right)$. Using the equivalence $\wedge^3 U \cong U^{^*}\otimes L_4^{^*}$, and employing the dual monad sequence twisted up with $L_4^{^*}$ and $L^{^*}$ in a similar fashion to the discussion in Appendix~\ref{sec:stab_criteria}, it follows that the relevant cohomology group can be expressed as: 
\beq
\begin{aligned}
H^0(L^{^*}\otimes \wedge^3 U) = &\ 
\text{Coker}\left(H^0\big( C^{^*}\!\otimes L_4^{^*}\otimes L^{^{\!*}} \big) \rightarrow H^0\big(B^{^*}\! \otimes L_4^{^*} \otimes L^{^{\!*}}\big)  \right)\\ &\oplus \text{Ker}\!\left(H^1\big( C^{^*}\!\otimes L_4^{^*}\otimes L^{^{\!*}} \big) \rightarrow H^1\big(B^{^*}\!\otimes L_4^{^*} \otimes L^{^{\!*}}\big)  \right)
\end{aligned}
\vspace{8pt}
\eeq

We have performed this computation for all the line bundles with entries between $-3$ and $3$, obtaining the following set of injecting line bundles: 
\begin{table}[!h]
\begin{center}
\begin{tabular}{ c c c c c c}
\varstr{14pt}{9pt} (3, 0, 0, 3),\, &(3, -1, 3, 2),\, &(3, -1, 2, 3),\, &(2, 0, 0, 3),\, &(2, -1, 3, 2),\, &(1, -1, 3, 2)
\end{tabular}
 \end{center}
 \vspace{-20pt}
\end{table}

As before, none of these line bundles de-stabilises the cone $C_{\widetilde{V}}$.
 }
 
\section{Conclusions and Outlook}
In this paper, we have presented an in-depth analysis of various aspects of heterotic line bundle models, in the context of a ``case-study" for the tetra-quadric Calabi-Yau hyper-surface in $(\IC\IP^1)^{\times 4}$. First, we have studied the question of finiteness of the class of heterotic line bundle models; mathematically this corresponds to the question of finiteness of poly-stable line bundles sums with fixed total Chern class. On the tetra-quadric, we have proved this finiteness result for the simple case of rank-two line bundle sums of the form $L\oplus L^{^*}$. For higher rank line bundle sums, we have seen that the complication arises at the boundary of K\"ahler moduli space. Indeed, restricting to the region in K\"ahler moduli space away from the boundaries (which corresponds to the supergravity approximation) and demanding a finite Calabi-Yau volume, we can show finiteness relatively easily.

From the $94$ phenomemologically promising models on the tetra-quadric which arise for the available group orders $|\Gamma|=2,4,8,16$ of freely-acting symmetries, we have chosen a particular example, exhibiting a symmetry $\Gamma=\mathbb{Z}_2\times\mathbb{Z}_2$, and we have used this for a more detailed discussion. At the GUT level, this model has gauge group $SU(5)\times U(1)^4$ (with three of the $U(1)$ anomalous) and twelve ${\bf 10}\oplus\overline{\bf 5}$ multiplets which lead to precisely three standard model families in the downstairs theory. There are three ${\bf 5}-\overline{\bf 5}$ pairs which, for appropriate Wilson line choices, lead to one pair of Higgs doublets downstairs. In addition, there are $60$ bundle moduli singlets, which reduce to $15$ downstairs. In summary, downstairs, this model has precisely the MSSM spectrum charged under the standard model group plus a number of standard model singlets.

For this model, we have studied the continuation into the non-Abelian part of the moduli space, using the four-dimensional effective theory as well as two different bundle constructions - extensions bundles and monads. Our particular focus was the fate of the Higgs doublets. We found that the only superpotential $\mu$-term allowed by the $U(1)^4$ symmetry is of the form ${\bf 1}_{2,4}\,L\,\bar{H}$, where ${\bf 1}_{2,4}$ is one type of bundle moduli singlets. Hence, in the part of bundle moduli space where $\langle {\bf 1}_{2,4}\rangle =0$ (which includes the line bundle locus, but also non-Abelian  $S\left(U(4)\times U_X(1)\right)$-bundles) the Higgs doublets should remain massless. We have verified this prediction by explicitly constructing the associated non-Abelian bundles, both through extensions and monads, and computing their cohomologies. Hence, we have found a model where the Higgs doublets remain massless away from the purely Abelian line bundle locus, although not in the whole bundle moduli space. We have shown that the remaining $U_X(1)$ gives rise to a $B-L$ symmetry of the model.

The line bundle data base~\cite{lbdatabase} contains some models, defined on more complicated Calabi-Yau manifolds, where all $\mu$-terms are forbidden by the $U(1)^4$ symmetry. It would be interesting to explicitly study the non-Abelian continuations of these models. We hope that the results of the present paper will be helpful in this context and, more generally, for the task of finding a realistic standard model from string theory.

\section*{Acknowledgements}
The work of  E.~I.~B.~is supported by the ARC Future Fellowship FT120100466.
A.~L.~is partially supported by the EPSRC network grant EP/l02784X/1 and by the STFC consolidated grant ST/L000474/1. A.~C.~wishes to thank the University College, Oxford and the STFC for supporting his graduate studies.

\newpage
\appendix

{\setstretch{1.2}
\section{The Tetraquadric Hypersurface}\label{app:KahlerCone}
Let $X$ denote a generic (smooth) hypersurface embedded in a product of four $\IC\IP^{1}$ spaces, $\cA=(\IC\IP^1)^{\times 4}$ defined as the zero locus of a polynomial of multi-degree (2,2,2,2) in the homogeneous coordinates of the four projective spaces:
\vspace{12pt}
\begin{equation}
X~=~~
\cicy{\IC\IP^1 \\   \IC\IP^1\\ \IC\IP^1\\ \IC\IP^1}
{ ~2 \!\!\!\!\\
  ~2\!\!\!\! & \\
  ~2\!\!\!\! & \\
  ~2\!\!\!\!}_{-128}^{4,68}\
\end{equation}
Let $\{J_i, 1\leq i\leq4\}$ denote the standard K\"ahler forms on the four embedding $\IC\IP^{1}$ spaces. Their restrictions, $\left.J_i\right|_X$, span $H^2(X, TX)$. In the following, we will use the same notation $J_i$ when referring to the restrictions of the K\"ahler forms to $X$. A K\"ahler form $J=t^iJ_i$ on $X$ is defined by a set of four real parameters $t^i$ which take values in the K\"ahler cone
\beq
C_{{\bf t}} =\left\{ {\bf t} \in \IR^4\ \left|\ t^i\geq 0,\, 1\leq i\leq 4\right. \right\} 
\eeq
The supergravity approximation of the heterotic string is valid for $t^i\gg 1$.

The second Chern class of $X$ is given by $c_2(X).J_i = (24,24,24,24)$ and the triple intersection numbers have the following simple form:
\beq
d_{ijk} = \int_X J_i\wedge J_j\wedge J_k = \begin{cases} 2 & \mbox{ if } i\neq j, j\neq k \\ 0 &\mbox{ otherwise } \end{cases}
\eeq

This leads to the following expressions for the volume $\kappa = d_{ijk}\, t^i\,t^j\,t^k$ and its first derivatives $\kappa_i = d_{ijk}\,t^j\,t^k$
\begin{align}
\kappa & = 12\, \left( t_1\,t_2\,t_3 + t_1\,t_2\,t_4 + t_1\,t_3\,t_4 + t_2\,t_3\,t_4\right)\\
\kappa_{i_1} & = 4\, \left( t_{i_2}\,t_{i_3} + t_{i_2}\,t_{i_4} +t_{i_3}\,t_{i_4} \right)\; , 
\end{align}
where $\left(i_1, i_2, i_3, i_4 \right)$ is any permutation of $(1,2,3,4)$. The derivatives $\kappa_i$ turn out to be useful for  discussing the slope $\mu_{\bf t}(V)$ of vector bundles $V$, defined as $\mu_{\bf t}(V) = d_{ijk}\, c_1^i(V)\, t^j\, t^k = c_1^i(V)\,\kappa_i$. 

In order to ensure that supersymmetry is preserved by the gauge fields, we have to require the vanishing of the slope. This condition may restrict the allowed values of the K\"ahler parameters to a subset of the K\"ahler cone. Since it is easier to discuss these constraints in the $\kappa_i$ variables, let us define the dual K\"ahler cone as the image of $C_{\bf t}$ under the map $f({\bf t}) = \left( \kappa_1, \kappa_2, \kappa_3, \kappa_4\right) /4$. Denote the new coordinates by ${\bf s}\in \IR^4$ and define the vectors ${\bf n}_i = {\bf n}-{\bf e}_i$, where ${\bf n} = (1,1,1,1)/2$ and ${\bf e}_i$ are the standard basis vectors in $\IR^4$. The dual K\"ahler cone, $f(C_{{\bf t}})$ is contained as a dense subset in the cone 
\beq 
C_{{\bf s}} = \left\{ {\bf s}\in\IR^4 \ \left|\  {\bf n}_i\cdot {\bf s} \geq 0,\ {\bf e}_i\cdot {\bf s} \geq 0,\ 1\leq i\leq 4\right.\right\}\; .
\eeq
The inclusion $f(C_{{\bf t}}) \subset C_{{\bf s}}$ is straightforward: since $t^i\geq 0$ for all $1\leq i\leq 4$, ${\bf e}_i\cdot f({\bf t}) = \kappa_i/4 \geq 0$ and ${\bf n}_i\cdot f({\bf t}) = \sum_{j\neq i} t_i\,t_j \geq 0$. In order to prove that the inclusion is dense, we need to show that almost every point ${\bf s}\in C_{{\bf s}}$ has at least one pre-image in $C_{\bf t}$. For this purpose, it is useful to define a new set of coordinates $x_i = {\bf n}_i\cdot{\bf s}$. Then ${\bf s}\in C_{\bf s}$ is equivalent with $x_i\geq 0$ and ${\bf n}_i\cdot {\bf x}\geq 0$, since ${\bf e}_i\cdot {\bf s}  =\left( {\bf n}-{\bf n}_i\right) \cdot {\bf s} = \left( \frac{1}{2}\sum{\bf n}_j-{\bf n}_i\right) \cdot {\bf s} = \frac{1}{2}\sum x_j - x_i = {\bf n}_i \cdot {\bf x}$. Without loss of generality, we will assume in the following that $x_1\leq x_2\leq x_3 \leq x_4$. 

If ${\bf t}$ is a pre-image of ${\bf x}$, then 
\beq 
{\bf n}_i\cdot f({\bf t})  = t_i \left(\sum_{j=1}^4 t_j - t_i \right) \stackrel{!}{=} x_i
\eeq

This can happen if we write 
\beq
t_i = \tau - \epsilon_i \sqrt{\tau^2 - x_i}\ \ \ \ \ \ \tau = \frac{1}{2} \sum_{i=1}^4 \epsilon_i \sqrt{\tau^2-x_i} 
\eeq
where $\epsilon_i$ are signs and we choose $\epsilon_1 = \epsilon_2 = \epsilon_3 =1$. Deciding whether ${\bf x}$ has a pre-image ${\bf}$ amounts to deciding whether the function $g: [\sqrt{x_4}, \infty) \rightarrow \IR$, defined by
\beq 
g(\tau)  = \tau - \frac{1}{2} \sum_{i=1}^4 \epsilon_i \sqrt{\tau^2-x_i}
\eeq
vanishes somewhere. Since $g$ is continuous, the existence of a vanishing point is guaranteed if $g\left(\sqrt{x_4}\right)$ and $g(\tau)_{\tau\rightarrow\infty}$ have different signs. We can use the freedom of choosing $\epsilon_4$ in order to arrange for this. Note, however, that $g\left(\sqrt{x_4}\right)$ does not depend on this choice.

\vspace{12pt}
If $g\left( \sqrt{x_4}\right)=0$, we already have a vanishing point.  This happens if $x_1=x_2=0$ and $x_3=x_4$. Otherwise, if $g\left( \sqrt{x_4}\right)>0$, we can choose $\epsilon_4=1$, such that $g(\tau)_{\tau\rightarrow\infty}<0$. If $g\left( \sqrt{x_4}\right)<0$ we choose $\epsilon_4 = -1$. However, this final case needs a more extended discussion. In this case, for large~$\tau$,
\beq 
g(\tau) \simeq \frac{1}{4\,\tau}\left(x_1 + x_2 + x_3 - x_4\right) + \frac{1}{16\,\tau^3}\left(x_1^2 + x_2^2 + x_3^2 - x_4^2\right)+\ldots
\eeq
If $x_1 + x_2 + x_3 - x_4>0$, i.e.~${\bf n}_4\cdot {\bf x}>0$, then $g(\tau)_{\tau\rightarrow\infty} >0$, which guarantees the existence of a vanishing point in the interval $[\sqrt{x_4}, \infty)$. If ${\bf n}_4\cdot {\bf x}=0$ and $x_2 > 0$ (the second condition excludes the case $x_1=x_2=0$ discussed above), $g(\tau)_{\tau\rightarrow\infty} < 0$, thus the above criterion cannot be used in order to decide whether $g$ has any vanishing points. In fact, in these cases it is easy to check that the equations $f({\bf t}) = {\bf s}$ are inconsistent. Thus the corresponding points, which lie on the boundary of $C_{\bf s}$ do not belong to the dual K\"ahler cone. 
}

\section{Topological Identities for Line Bundles on the Tetraquadric}\label{app:lbtopology}

A line bundle $L$ is determined by its first Chern class, $c_1(L) = k^i J_i$ and it is also denoted by $L = \cO_X({\bf k})$. The Chern character  for a single line bundle is given by
\beq
\text{ch}(L) = e^{\,c_1(L)}  = 1 + c_1(L) + \frac{1}{2} c_1(L)^2+ \frac{1}{3!} c_1(L)^3 
\eeq

This implies, on the tetraquadric, 
\begin{align}
\text{ch}_2(L).J_i &= \int_X \frac{1}{2} c_1(L)^2 \wedge J_i = \frac{1}{2}\,d_{ijk}\,k^j\,k^k = 2\sum_{\stackrel{j<k}{j,k\neq i}} k^j\,k^k\\
\int_X \text{ch}_3(L) & = \frac{1}{3!}\, d_{ijk}\,k^i\,k^j\,k^k = 2\sum_{i<j<k} k^i\,k^j\,k^k
\end{align}

For a sum of line bundles, $\displaystyle V =\bigoplus_{a=1}^r \cO_X({\bf k}_a)$, given the additivity of the Chern characters, we have
\begin{align}
\text{ch}_1(V) &= \left(\sum_{a=1}^r k^i_a \right) J_i\\
\label{eq:c2}\text{ch}_2(V).J_i &= \frac{1}{2}\,d_{ijk}\sum_{a=1}^r\,k_a^i\,k_a^j=  2\sum_a\sum_{\stackrel{j<k}{j,k\neq i}} k_a^j\,k_a^k\\
\int_X \text{ch}_3(V) & = \frac{1}{3!}\, d_{ijk}\sum_{a=1}^r\,k_a^i\,k_a^j\,k_a^k = 2\sum_{a=1}^r\sum_{i<j<k} k^i\,k^j\,k^k
\end{align}

The index of $V$ on a Calabi-Yau manifold $X$, for which $\text{Td}(TX) = 1 +c_2(TX)/12$ can be computed using the index theorem:
\beq
\begin{aligned} 
\text{ind} (V) & = \int_X \text{ch}(V)\, \text{Td}(TX) = \int_X \text{ch}_3(V) + \frac{1}{12}\, \text{ch}_1(V)\,c_2(TX)\\
& = \frac{1}{6} \sum_{a=1}^r \left( d_{ijk}\,k_a^i\,k_a^j\,k_a^k +\frac{1}{2}\, k_a^i \left(c_2(TX).J_i \right) \right) \\
& = 2\sum_{a=1}^r\sum_{i<j<k} k^i\,k^j\,k^k + 2\sum_{a=1}^r \sum_{i} k_a^i
\end{aligned}
\eeq

The slope of a line bundle is defined as
\beq 
\mu_{\bf t}(L) = \int_X c_1(L) \wedge J\wedge J = d_{ijk}\,k^i\,t^j\,t^k = k^i\kappa_i
\eeq
and will play a central role in deciding stability of vector bundles.

\section{Line Bundle Cohomology on the Tetraquadric}\label{app:tqcoh}
\vspace{-8pt}
Let $X$ denote the tetraquadric hypersurface. If the defining polynomial is sufficiently generic, $X$ is a smooth Calabi-Yau manifold with $h^{1,1}(X)=4$. Moreover, this embedding is favourable, in the sense of \cite{Anderson:2013xka}. This implies that the second cohomology of $X$ descends entirely from that of the embedding space. As such, all possible line bundles on $X$ can be obtained as restrictions  to $X$ of line bundles on~$\cA$. Let $L={\cal O}_X({\bf k})$ be a line bundle with first Chern class defined by a set of four integers $k^i$ as $c_1(L)=k^iJ_i$. For the ease of notation, we order these line bundle integers such that $k_1\leq k_2 \leq k_3\leq k_4$ in the following discussion.

\vspace{12pt}
The ranks of the ambient space cohomology groups can be obtained using Bott's formula:
\beq\label{bott}
h^q(\IC\IP^n,\cO(k)) =
\left\{\ba{lll}
~~{k+n \choose n} & q = 0 & k\geq 0,\\[12pt]
{-k-1 \choose -k-n-1} & q = n & k\leq-n-1,\\[9pt]
~~~~~0 & \mbox{otherwise} & 
\ea\right.
\eeq
and  the K\"unneth formula, which gives the cohomology of line bundles over direct products of
projective spaces $\cA = \IC\IP^{n_1} \times \ldots \times \IC\IP^{n_m}$:
\beq\label{kunneth1}
H^q(\cA, \cO\big(k_1, \ldots, k_m)\big) =
\bigoplus_{q_1+\ldots+q_m = q} H^{q_1}\big(\IC\IP^{n_1},\cO(k_1)\big) \times
\ldots \times H^{q_m}\big(\IC\IP^{n_m},\cO(k_m)\big) \ ,
\eeq

\vspace{10pt}
Thus, for a product of four projective spaces $\cA=(\IC\IP^1)^{\times 4}$, we obtain: 
\begin{align}
\mathrm{dim}\, H^0\big{(}\cA, \calO ({\bf k}) \big{)} & = \prod_{i=1}^4 (k_i +1)^+\\
\mathrm{dim}\, H^1\big{(}\cA, \calO ({\bf k}) \big{)} & =  (-k_1 -1)^+ \prod_{i=2}^4 (k_i +1)^+ \\[4pt]
\mathrm{dim}\, H^2\big{(}\cA, \calO ({\bf k}) \big{)} & =  (-k_1 -1)^+ (-k_2 -1)^+ (k_3 +1)^+(k_4 +1)^+\\[4pt]
\mathrm{dim}\, H^3\big{(}\cA, \calO ({\bf k}) \big{)} & = (k_4 +1)^+\prod_{i=1}^3(-k_i -1)^+ \\
\mathrm{dim}\, H^4\big{(}\cA, \calO ({\bf k}) \big{)} & = \prod_{i=1}^4 (-k_i -1)^+
\end{align}
where the notation $f^+$ indicates the positive part of a function $f^+(x)= \mathrm{max}\,(f(x),0)$.
 
\vspace{21pt}
Line bundle cohomology on $X$ can be computed using the Koszul resolution sequence
\beq
0\ \longrightarrow\ \cO_{\cA}({\bf k}) \otimes \cN_X^{^{*}} \longrightarrow\ \cO_{\cA}({\bf k})\ \longrightarrow\ \cO_{X}({\bf k})\ \longrightarrow\ 0
\eeq
where $\cN_X^{^{*}}$ is the dual to the normal bundle $\cN_X = \cO_{\cA}(2,2,2,2)$. The associated long exact sequence in cohomology reads: 
\begin{equation}
\begin{array}{ccccccccc}
0& \longrightarrow & H^0(\cA,\cO({\bf k}) \otimes \cN_X^{^{*}})& \longrightarrow& H^0(\cA,\cO({\bf k})) & \longrightarrow&H^0(X,\cO({\bf k}))&\longrightarrow&\\[4pt]
& \longrightarrow & H^1(\cA,\cO({\bf k}) \otimes \cN_X^{^{*}})& \longrightarrow& H^1(\cA,\cO({\bf k})) & \longrightarrow&H^1(X,\cO({\bf k}))&\longrightarrow&\\[4pt]
& \longrightarrow & H^2(\cA,\cO({\bf k}) \otimes \cN_X^{^{*}})& \longrightarrow& H^2(\cA,\cO({\bf k})) & \longrightarrow&H^2(X,\cO({\bf k}))&\longrightarrow&\\[4pt]
& \longrightarrow & H^3(\cA,\cO({\bf k}) \otimes \cN_X^{^{*}})& \longrightarrow& H^3(\cA,\cO({\bf k})) & \longrightarrow&H^3(X,\cO({\bf k}))&\longrightarrow&0\\
\end{array}
\end{equation}
Thus we can express the cohomology groups on $X$ as
\beq
\begin{aligned}
H^{q}\big(X,\cO({\bf k})\big)\  =\ \ & \text{Coker}\left( H^q(\cA,\cO({\bf k}) \otimes \cN_X^{^{*}}) \longrightarrow H^q(\cA,\cO({\bf k}))  \right) \ \oplus\\
\oplus\ \ & \text{Ker}\left( H^{q+1}(\cA,\cO({\bf k})\otimes \cN_X^{^{*}}) \longrightarrow H^{q+1}(\cA,\cO({\bf k}))  \right)
\end{aligned}
\eeq
In general, the ranks of these maps are maximal, in which case 
\begin{equation}
\begin{aligned}
\mathrm{dim}\, H^0\big{(}X, \calO ({\bf k}) \big{)}  & = \Big( \prod_{a=1}^4 (k_a +1)^+  - \prod_{a=1}^4 (k_a -1)^+\Big)^ + \\
& + \Big((-k_1 +1)^+ \prod_{a=2}^4 (k_a -1)^+ -(-k_1 -1)^+ \prod_{a=2}^4 (k_a +1)^+ \Big)^ + \\[7pt]
\mathrm{dim}\, H^1 \big{(}X, \calO ({\bf k}) \big{)}  & = \Big((-k_1 -1)^+ \prod_{a=2}^4  (k_a +1)^+- (-k_1 +1)^+ \prod_{a=2}^4 (k_a -1)^+  \Big)^ + \\
 & + \Big( \prod_{a=1}^2(-k_a+1)^+\prod_{a=3}^4(k_a-1)^+ -  \prod_{a=1}^2(-k_a-1)^+\prod_{a=3}^4(k_a+1)^+ \Big)^+ \\[7pt]
\mathrm{dim}\, H^2\big{(}X, \calO ({\bf k}) \big{)}  & =  \Big( \prod_{a=1}^2(-k_a-1)^+\prod_{a=3}^4(k_a+1)^+ -  \prod_{a=1}^2(-k_a+1)^+\prod_{a=3}^4(k_a-1)^+ \Big)^+ \\
& + \Big( (k_4 -1)^+\prod_{a=1}^3(-k_a + 1)^+ - (k_4 +1)^+\prod_{a=1}^3(-k_a -1)^+  \Big)^+ \\[7pt]
\mathrm{dim}\, H^3\big{(}X, \calO ({\bf k}) \big{)} &  = \Big( (k_4 +1)^+\prod_{a=1}^3(-k_a -1)^+ - (k_4 -1)^+\prod_{a=1}^3(-k_a + 1)^+ \Big)^+ \\
& - \Big( \prod_{a=1}^4 (-k_a +1)^+- \prod_{a=1}^4 (-k_a -1)^+ \Big)^+
\end{aligned}
\end{equation}

However, even in the cases when these ranks are non-maximal we are able to write down closed-form expressions for the ranks of the cohomology groups on $X$. This is due to the fact that the ranks of the maps involved exhibit a surprising regularity.  These formulae are given below and hold for any non-negative integer $p$. We use the notation $C_n^k$ for the binomial coefficient indexed by $n$ and $k$. We distinguish the following cases:
\vspace{2pt}
\begin{itemize}
\item[$i)$] $k_1\leq -(4+2p)$, $k_2= k_3 = -(4+2p)$ and $k_4=2+p$:
\end{itemize}
\begin{equation}
\begin{aligned}
\mathrm{dim}\, H^0\big{(} X, \calO ({\bf k}) \big{)}  &= \mathrm{dim}\, H^1\big{(}X, \calO ({\bf k}) \big{)}  = 0 \\[8pt]
\mathrm{dim}\, H^2\big{(}X, \calO ({\bf k}) \big{)}  & = 48\,C^{3}_{p+3} + (p+1)\,(-k_1-(4+2p)-1)\\
\mathrm{dim}\, H^3\big{(}X, \calO ({\bf k}) \big{)}  & = (k_4 +1)^+\prod_{a=1}^3(-k_a -1)^+ \\
-\, \Big((k_4 -1)^+ &\prod_{a=1}^3(-k_a + 1)^+ - \big(48\,C^{3}_{p+3} + (p+1)\,(-k_1-(4+2p)-1) \big)  \Big)
\end{aligned}
\end{equation}

\vspace{4pt}
\begin{itemize}
\item[$ii)$] $k_1, k_2< -(4+2p)$, $k_3\leq-(4+2p)$ and $k_4=2+p$ or \\[4pt] $k_1,k_2< -(4+2p)-2$, $k_3 = -(4+2p)+1$ and $k_4=2+p$:
\end{itemize}
\begin{equation}
\begin{aligned}
\mathrm{dim}\, H^0\big{(}X, \calO ({\bf k}) \big{)}  &= \mathrm{dim}\, H^1\big{(}X, \calO ({\bf k}) \big{)}  = 0 \\[8pt]
\mathrm{dim}\, H^2\big{(}X, \calO ({\bf k}) \big{)}  &= 48\,C^{3}_{p+3}\\
\mathrm{dim}\, H^3\big{(}X, \calO ({\bf k}) \big{)}  & = (k_4 +1)^+\prod_{a=1}^3(-k_a -1)^+ - \Big((k_4 -1)^+\prod_{a=1}^3(-k_a + 1)^+ -48\,C^{3}_{p+3}  \Big)
\end{aligned}
\end{equation}

\vspace{8pt}
The last two cases represent the dual version of first two. Although very similar, we include these formulae here for the sake of completeness:
\begin{itemize}
\item[$iii)$] $k_1=-2-p$, $k_2=k_3=4+2p$, $k_4\geq 4+2p$:
\end{itemize}
\begin{equation}
\begin{aligned}
\mathrm{dim}\, H^0\big{(}X, \calO ({\bf k}) \big{)}  & = (-k_1 +1)^+ \prod_{a=2}^4 (k_a -1)^+\\
 - \Big((-k_1  - &1)^+ \prod_{a=2}^4 (k_a +1)^+ - \big(48\,C^{3}_{p+3} + (p+1)\,(k_4-(4+2p)-1) \big)  \Big)\\[8pt]
\mathrm{dim}\, H^1\big{(}X, \calO ({\bf k}) \big{)}  & = 48\,C^{3}_{p+3} + (p+1)\,(k_4-(4+2p)-1)\\[8pt]
\mathrm{dim}\, H^2\big{(}X, \calO ({\bf k}) \big{)}  &= \mathrm{dim}\, H^3\big{(} \calO_X ({\bf k}) \big{)} = 0
\end{aligned}
\end{equation}

\vspace{8pt}
\begin{itemize}
\item[$iv)$] $k_1=-2-p$, $k_2\geq 4+2p$, $k_3, k_4> 4+2p$ or \\[4pt] $k_1=-2-p$, $k_2=4+2p-1$, $k_3, k_4> 4+2p+2$:
\end{itemize}
\begin{equation}
\begin{aligned}
\mathrm{dim}\, H^0\big{(} \calO_X ({\bf k}) \big{)}  &= (-k_1 +1)^+ \prod_{a=2}^4 (k_a -1)^+ - \Big((-k_1 -1)^+ \prod_{a=2}^4 (k_a +1)^+ - 48\,C^{3}_{p+3}  \Big)\\
\mathrm{dim}\, H^1\big{(} \calO_X ({\bf k}) \big{)}  &= 48\,C^{3}_{p+3}\\[8pt]
\mathrm{dim}\, H^2\big{(} \calO_X ({\bf k}) \big{)}  &=  \mathrm{dim}\, H^3\big{(} \calO_X ({\bf k}) \big{)} = 0
\end{aligned}
\end{equation}

\section{Finiteness: a Different Perspective}\label{app:finiteness}

In Section~\ref{sec:finiteness} we have tried to understand the observation, made in Ref.~\cite{Anderson:2013xka}, that the number of consistent and physically viable line bundle models constructed over a certain manifold $X$ admitting discrete symmetries $\Gamma$ of a fixed order is an increasing and saturating function of the maximal line bundle entry in modulus. This phenomenon could be observed for all the pairs $(X,|\Gamma|)$ studied in Ref.~\cite{Anderson:2013xka}. In this section we will argue that the combined effect of cohomology constraints, poly-stability and the bound on the second Chern class imposed by the anomaly cancellation condition limits the range of allowed line bundle integers.

 \subsection{Bounds from cohomology}
 
In order to ensure the correct chiral asymmetry and the absence of $\overline{\bf{10}}$-multiplets form the $SU(5)$ GUT spectrum, we must require the following pattern for the bundle cohomology: 
\beq
 h^{^{\!\bullet}}(X,V)\ \,  =\  (0,3\,|\Gamma|,0,0)  
\eeq

\vspace{0pt}
In Appendix~\ref{app:tqcoh} we have seen explicit formulae for line bundle cohomology on the tetraquadric hypersurface. By using these, and imposing the above cohomology pattern we can immediately exclude the following line bundles (where we assume $k_1\leq k_2\leq k_3\leq k_4$):
\begin{itemize}
\item[-] the trivial line bundle $\calO_X (0,0,0,0)$;
\item[-] all semipositive line bundles $\calO_X (k_1,k_2,k_3,k_4)$ with $k_i\geq 0$;
\item[-] $\calO_X (-p,0,0,k_4)$ with $p\geq 1$ and $k_4\geq 2$;
\item[-] $\calO_X (-1,k_2,k_3,k_4)$ with $k_2,k_3,k_4\geq 2$;
\item[-] all special cohomology cases presented in Section \ref{app:tqcoh}
\item[-] $\calO_X (k_1,k_2, k_3,k_4)$ with $k_1,k_2<0$ and $k_3, k_4=0$
\item[-] $\calO_X (k_1,k_2,k_3,k_4)$ with $k_1,k_2,k_3<0$ and $k_4\geq0$
\end{itemize}

Thus we are left with the following classes of line bundles:
\begin{itemize}
\item[-] $\calO_X (-p,0,0,1)$ with $p\geq 1$
\item[-] $\calO_X (-p,0,k_3,k_4)$ with $k_3,k_4>0$ 
\item[-] $\calO_X (-p,k_2,k_3,k_4)$ with $k_2<4+2p-1$ 
\item[-] $\calO_X (-p,k_2,k_3,k_4)$ with $k_2=4+2p-1$ and $k_3<4+2p+2$ 
\item[-] $\calO_X (k_1,k_2,k_3,k_4)$ with $k_1,k_2<0$ and $k_3,k_4>0$ 
\end{itemize} 

\vspace{20pt}
Moreover, for any line bundle $L$ in a viable line bundle sum, we must require $h^1\big{(} X,L \big{)} \leq 3| \Gamma |$, where $\Gamma$ is a freely acting group on the tetraquadric hypersurface. According to Ref.~\cite{Braun:2010vc}, the available group orders are $| \Gamma | \in \{2,4,8,16\}$, so that  $h^1\big{(} X,L\big{)} \leq 48$. This imposes further bounds on the allowed line bundle entries. It follows that the allowed line bundles (assuming $k_1\leq k_2\leq k_3\leq k_4$) fall into several categories: 
\begin{itemize}
\item[$i)$] one negative and one positive entry:
	\begin{itemize}
	\item[-] $\calO_X ([-25,-1],0,0,1)$. Cohomology: $h^{^{\!\bullet}} \big{(} \calO_X (-p,0,0,1) \big	{)} = \big(0,2(p-1),0,0\big)$
	\end{itemize}
	
\item[$ii)$] one negative and two positive entries:
	\begin{itemize}
	\item[-] $\calO_X (-1,0,1,[1,\infty))$. Cohomology: $h^{^{\!\bullet}} \big{(} \calO_X (-1,0,1,q) \big{)} = \big(0,0,0,0\big)$
	\item[-] $\calO_X (-k_1,0,k_3,k_4)$, $n_i\in \mathbb Z_+$. Cohomology: $h^{^{\!\bullet}} \big{(} \calO_X (-k_1,0,k_3,k_4) \big{)} =$\\ $ \big(0,(-k_1 -1)^+ (k_3 +1)^+(k_4 +1)^+ + (-k_1 +1)^+ (k_3 -1)^+(k_4 -1)^+ ,0,0\big)$. \\Bound on the entries: $k_1, k_3, k_4 \leq 25$
	\end{itemize}
	
\item[$iii)$] one negative and three positive entries:
	\begin{itemize}
	\item[-] $\calO_X ([-7,-1],1,1,1)$. Cohomology: $h^{^{\!\bullet}} \big{(} \calO_X (-p,1,1,1) \big{)} = \big(0,8(p-1),0,0\big)$
	\item[-] $\calO_X (-1,1,[1,\infty),[1,\infty))$. Cohomology: $h^{^{\!\bullet}} \big{(} \calO_X (-1,1,p,q) \big{)} = \big(0,0,0,0\big)$
	\item[-] $\calO_X (-p,k_2,k_3,k_4)$, $p>1,k_i\in\mathbb Z_+$.  Bound on the entries: for $p=2$, $k_i\leq 11$; for $p=3$, $k_i\leq 5$; for $p=4$, $k_i\leq 3$; for $p=5$, $k_i\leq 2$; for $p=6, 7$, $k_i = 1$; 
	\end{itemize}

\item[$iv)$] two negative entries and one positive entry:
	\begin{itemize}
	\item[-] $\calO_X ((-\infty,-1],-1,0,1)$. Cohomology: $h^{^{\!\bullet}} \big{(} \calO_X (-p,-1,0,1) \big{)} = \big(0,0,0,0\big)$
	\end{itemize}
	
\item[$v)$] two negative and two positive entries:
\vspace{-8pt}
	\begin{itemize}
	\item[-] $\calO_X (-p,-q,p,q)$, $p,q\in \mathbb Z_+$. Cohomology: $h^{^{\!\bullet}} \big{(} \calO_X (-p,-q,p,q) \big{)} = \big(0,0,0,0\big)$
	\item[-] $\calO_X ((-\infty,-1],(-\infty,-1],[1,\infty),[1,\infty))$. Cohomology: $h^{^{\!\bullet}} \big{(} \calO_X (-p,-q,q,p) \big{)} = \big(0,0,0,0\big)$
	\item[-] $\calO_X (-k_1,-k_2,k_3,k_4)$, $k_i\in\mathbb Z_+$. Cohomology: $h^{^{\!\bullet}} \big{(} \calO_X (-k_1,-k_2,k_3,k_4) \big{)} =$\\ $ \big(0,(-k_1 +1)^+ (-k_2 +1)^+ (k_3 -1)^+(k_4 -1)^+ -(-k_1 -1)^+ (-k_2 -1)^+ (k_3 +1)^+(k_4 +1)^+,0,0\big) $ \newline
	Bound on the entries (for cases not covered above): $n_i\leq 13$
	\end{itemize}
	
\item[$vi)$] three negative entries and one positive entry:
\vspace{-8pt}
	\begin{itemize}
	\item[-] $\calO_X ((-\infty,-1],(-\infty,-1],-1,1)$. Cohomology: $h^{^{\!\bullet}} \big{(} \calO_X (-p,-q,-1,1) \big{)} = \big(0,0,0,0\big)$
	\end{itemize}
\end{itemize}
 
\vspace{20pt} 
\subsection{A bound from stability}\label{sec:boundstab}
Poly-stability for a sum of five line bundles $V = \bigoplus_{a=1}^5 L_a$ reduces to the question of finding simultaneous solutions for the equations $\mu_{\bf t}(L_a) = 0$, that is finding a vector $(\kappa_1,\kappa_2,\kappa_3,\kappa_4)$ in the dual K\"ahler cone, such that for all $1\leq a \leq 5$,
\beq
0=\mu_{\kappa}(L_a) =\kappa_1 k_a^1 +\kappa_2 k_a^2 +\kappa_3 k_a^3 +\kappa_4 k_a^4 \; .
\eeq
However, due to the condition $c_1(V)=0$, only four of these equations are independent. By successively multiplying the above equation by $k_a^1,\ldots, k_a^4$, we obtain
\begin{equation}
\begin{aligned}
0 & = \kappa_1 \left( k_a^1 \right)^2 +\kappa_2 k_a^2 k_a^1+\kappa_3 k_a^3 k_a^1+\kappa_4 k_a^4k_a^1 \\[4pt]
0 & = \kappa_1 k_a^1k_a^2 +\kappa_2  \left( k_a^2 \right)^2 +\kappa_3 k_a^3k_a^2 +\kappa_4 k_a^4 k_a^2 \\[4pt]
0 & = \kappa_1 k_a^1k_a^3 +\kappa_2 k_a^2 k_a^3+\kappa_3  \left( k_a^3 \right)^2 +\kappa_4 k_a^4k_a^3 \\[4pt]
0 & = \kappa_1 k_a^1 k_a^4+\kappa_2 k_a^2k_a^4 +\kappa_3 k_a^3 k_a^4+\kappa_4  \left( k_a^4 \right)^2 
\end{aligned}
\end{equation}
Adding these up and summing over  $a$, we obtain
\beq
\begin{aligned} 
0& = \kappa_1\left( \sum_{a=1}^5 \left( k_a^1 \right)^2 + \frac{-c_2^1+c_2^2+c_2^3+c_2^4}{2} \right)   + \kappa_2\left( \sum_{a=1}^5 \left( k_a^2 \right)^2 + \frac{+c_2^1-c_2^2+c_2^3+c_2^4}{2} \right)  \\
& + \kappa_3\left( \sum_{a=1}^5 \left( k_a^3 \right)^2 + \frac{+c_2^1+c_2^2-c_2^3+c_2^4}{2} \right)   + \kappa_4\left( \sum_{a=1}^5 \left( k_a^4 \right)^2 + \frac{+c_2^1+c_2^2+c_2^3-c_2^4}{2} \right)
\end{aligned}
\eeq
where
$c_2 ^1 :=   c_2 (V) . J_1 = \sum_{a=1}^5  k_a^2 k_a^3 +k_a^2 k_a^4 + k_a^3 k_a^4$
and analogously for the other indices.

\vspace{12pt}
Since $\kappa_i$ are all positive, and $c_2^i$ are bounded from above by the anomaly cancellation condition~\eqref{tqanom}, it follows that $ \sum_a \left( k_a^1 \right)^2$, $ \sum_a \left( k_a^2 \right)^2$, $ \sum_a \left( k_a^3 \right)^2$ and $ \sum_a \left( k_a^4 \right)^2$ cannot be large at the same time. That is, at least one of the rows in the matrix representing the sum of line bundles has to contain only small numbers. 

More significantly, in order to satisfy the above equation with large line bundle entries, some (but not all) of the $\kappa_i$ parameters have to become small. This corresponds to having some, but not all of the K\"ahler parameters $t^i$ arbitrarily small. Though not directly relevant for the present discussion, we mention that such a situation falls out of the supergravity approximation, valid when all the K\"ahler parameters satisfy $t^i\gg 1$. 

\subsection{Combined bounds}
The physical constraints on the cohomologies of $V$ limit the range of line bundle integers in almost all cases. However, these constraints leave three types of line bundles with large integers in modulus, having trivial cohomology. Up to permutations, these are:
\begin{itemize}
\item[-] line bundles with one large integer: $\cO_X(-1,0,1, \pm p)$
\item[-] line bundles with two large integers: $\cO_X(-1,1,\pm q, \pm p)$
\item[-] line bundles with four large integers: $\cO_X(-p,-q,q,p)$
\end{itemize}

The third type of line bundles are excluded by stability and the $c_2(V)$ constraint discussed above, in Section \ref{sec:boundstab}. Moreover, line bundles of the form $\cO_X(-1,1, p, q)$ where both $p$ and $q$ are either positive or negative do not satisfy the slope zero condition in the positive K\"ahler cone. Thus the only allowed type of line bundles with large entries are of two types: 
\begin{itemize}
\item[-] two large integers: $\cO_X(p,-q,-1, 1)$, $p, q\geq 0$ or 
\item[-] $\cO_X(-p,-q,q,p)$ where only $p\geq 0$ can be arbitrarily large
\end{itemize}

\vspace{21pt}
We would like to argue that such line bundles, with arbitrarily large entries, cannot enter a viable line bundle model. Before doing that, however, we give an example of an infinite family of line bundle sums satisfying all the topological constraints required from a viable model:  
\begin{equation}
V~=~~
\cicy{ \\ \\ \\ \\ }
{ - p & p-2 & ~~0 & ~~1 & ~~1~ \\
~~0 & ~~0 & -1 & ~~2 & -1~ \\
-1 & -1 & -1 & ~~1 &~~2 ~\\
 ~~1 & ~~1 &~~ 1 & -2 & -1 ~\\}\; 
\end{equation}

The bundles in this class satisfy, independent of the value of $p\in\IZ$:
\vspace{-4pt}
\begin{align}
c_1(V) &= 0\\  
c_2(V).J_i &= \left(-20, -14, -16, 14\right) \\
\text{ind}(V) &= \text{ind}(\wedge^2 V) = 12
\end{align}
\vspace{-21pt}

However, the slopes of the above line bundles cannot be simultaneously put to zero in the interior of the K\"ahler cone, except when $p=1$. 

\vspace{21pt}
The following family contains poly-stable line bundle sums for any $p\in 3\IZ$:
\begin{equation}
V~=~~
\cicy{ \\ \\ \\ \\ }
{ - p & p/3 & p/3 & p/3 & ~~0~ \\
 ~~0 & -1 & -1 & -1 & ~~3~ \\
-1 & ~~0 & ~~0 & ~~0 &~~1 ~\\
 ~~1 & ~~1 &~~ 1 & -2 & -4 ~\\}\; 
\end{equation}
and has 
\vspace{-20pt}
\begin{align}
c_1(V) &= 0\\  
\text{ind}(V) &= \text{ind}(\wedge^2 V) = 24
\end{align}

However, the second and the third entries in $c_2(V).J_i$ depend linearly on $p$.
The two examples above illustrate the conflict between, on one hand, trying to satisfy the slope zero condition and on the other hand trying to keep $c_2(V).J_i$ bounded. If one starts with a line bundle of the type $\cO_X(-p,-q,q,p)$, one is forced by the slope zero condition to continue with line bundles of the same kind. However, this always leaves at least two entries in $c_2(V).J_i$ large. Alternatively, if one starts with a line bundle $\cO_X(-p,q,-1,1)$, this will produce at least one large entry in $c_2(V).J_i$. Trying to make up for that renders the slope zero condition impossible to satisfy.

\section{Criteria for Stability of Monad Bundles}\label{sec:stab_criteria}
If a line bundle $L$ is a proper sub-bundle of $\wedge^k V$, then 
\begin{equation}
\text{Hom}_X\,\big( L,\wedge^k V\big)\cong H^0 \big( X, \wedge ^k V\otimes L^{^{\!*}} \big) \neq 0
\end{equation}

Thus $\text{Hom}_X\big(L,\wedge^k(V )\big) = 0$ guarantees that there are no maps between $L$ and $V$. Below, we derive several criteria for deciding whether $\text{Hom}_X\big(L,\wedge^k(V )\big)$ is trivial or not. We will use the fact that, for $SU(n)$ bundles, there exists the isomorphism
\beq
\wedge^{n-k}V \cong \wedge^k V^{^*}
\eeq

\vspace{2pt}
1. The case $k=1$, $L\hookrightarrow V$. By twisting the monad sequence (\ref{eq:monad}) with the line bundle $L^{^{\!*}}$, we obtain a short exact sequence which yields the following long exact sequence in cohomology:
\beq
0\ \longrightarrow\ \text{Hom}_X\big(L,V\big) \longrightarrow\ H^0\big(B\otimes L^{^{\!*}}\big)\ \longrightarrow\ H^0\big(C\otimes L^{^{\!*}}\big)\ \longrightarrow\ \ldots 
\vspace{-8pt}
\eeq
Thus 
\vspace{-8pt}
\beq\label{eq:hom}
 \text{Hom}_X\big(L,V\big) \cong \text{Ker}\left( H^0\big(B\otimes L^{^{\!*}}\big)\ \longrightarrow\ H^0\big(C\otimes L^{^{\!*}}\big)\right)
\eeq

If $h^0\big(B\otimes L^{^{\!*}}\big) = 0$, then $L$ does not inject into $V$. If 
\vspace{-4pt}
\beq\label{eq:cohcondiiton1}
h^0\big(B\otimes L^{^{\!*}}\big)> h^0\big(C\otimes L^{^{\!*}}\big)
\vspace{-4pt}
\eeq 
then $\text{Hom}_X\big(L,V\big)$ is non-trivial. In this case, we will treat $L$ as a de-stabilising line bundle and restrict the K\"ahler cone accordingly.  In the cases in which these simple cohomology checks are not conclusive, we will need to explicitly find the kernel of the map (\ref{eq:hom}).

\vspace{12pt}
2. The case $k=2$, $L\hookrightarrow \wedge^2 V$. Start with the second exterior power sequence:
\vspace{-4pt}
\beq\label{eq:sequence2}
0\ \longrightarrow\ \wedge^2V \longrightarrow\ \wedge^2B\ \longrightarrow\ B\otimes C\ \longrightarrow\ S^2C\ \longrightarrow\ 0
\vspace{-4pt}
\eeq
then twist this with $L^{^{\!*}}$ and split the resulting sequence into two short exact sequences. The associated long exact sequences in cohomology start as:
\begin{align}
&0\ \longrightarrow\ \text{Hom}_X\big(L,\wedge^2V\big) \longrightarrow\ H^0\big(\wedge^2B\otimes L^{^{\!*}}\big)\ \longrightarrow\ H^0\big(Q)\ \longrightarrow\ \ldots \\
&0\ \longrightarrow\ H^0\big(Q)\ \longrightarrow\ H^0\big(B\otimes C \otimes L^{^{\!*}}\big) \longrightarrow\ H^0\big(S^2C\otimes L^{^{\!*}}\big)\ \longrightarrow\ \ldots
\end{align}
Thus 
\vspace{-20pt}
\begin{align}\label{eq:hom2}
 \text{Hom}_X\big(L,\wedge^2V\big)& \cong \text{Ker}\left( H^0\big(\wedge^2B\otimes L^{^{\!*}}\big)\ \longrightarrow\ H^0\big(Q\big)\right)\\
 H^0(Q)&\cong \text{Ker}\left( H^0\big(B\otimes C\otimes L^{^{\!*}}\big)\ \longrightarrow\ H^0\big(S^2C\otimes L^{^{\!*}}\big)\right)
\end{align}

As before, there are two simple cohomology checks that can be performed in this case. If $h^0\big(\wedge^2B\otimes L^{^{\!*}}\big)=0$, then $L$ does not inject into $\wedge^2 V$. However, $\text{Hom}_X\big(L,\wedge^2V\big)$ is non-trivial, if 
\vspace{-8pt}
\beq\label{eq:cohcondiiton2}
h^0\big(\wedge^2B\otimes L^{^{\!*}}\big)> h^0\big(B\otimes C\otimes L^{^{\!*}}\big)
\eeq

3. The case $k=3$, $L\hookrightarrow \wedge^3 V\cong \wedge^2 V^{^*}$. There are two relevant sequences in this case. First, start with the dual sequence for $\wedge^2V$ twisted up with $L^{^{\!*}}$:
\vspace{-4pt}
\beq\label{eq:sequence3}
0\ \longrightarrow\ S^2C^{^*}\otimes L^{^{\!*}} \longrightarrow\ B^{^*}\otimes C^{^*} \otimes L^{^{\!*}}\ \longrightarrow\ \wedge^2 B^{^*} \otimes L^{^{\!*}}\ \longrightarrow\ \wedge^2V\otimes L^{^{\!*}}\ \longrightarrow\ 0
\vspace{-4pt}
\eeq
and split this into two short exact sequences with associated long exact sequences in cohomology:
\vspace{-4pt}
\begin{align*}
&0\ \longrightarrow\ H^0\big( S^2C^{^*}\otimes L^{^{\!*}} \big) \longrightarrow\ H^0\big(B^{^*}\otimes C^{^*} \otimes L^{^{\!*}}\big)\ \longrightarrow\ H^0\big(Q)\ \longrightarrow\ H^1\big( S^2C^{^*}\otimes L^{^{\!*}} \big) \longrightarrow\ \ldots \\
&0\ \longrightarrow\ H^0\big(Q)\ \longrightarrow\ H^0\big(\wedge^2 B^{^*} \otimes L^{^{\!*}}\big) \longrightarrow\ \text{Hom}_X\big(L,\wedge^2V^{^*}\big)\ \longrightarrow\ H^1\big(Q)\ \longrightarrow\ \ldots
\vspace{-4pt}
\end{align*}

\vspace{-8pt}
From here we can infer that $\text{Hom}_X\big(L,\wedge^3V\big) = 0$ if $h^0\big(\wedge^2 B^{^*} \otimes L^{^{\!*}}\big) = h^1\big( S^2C^{^*}\otimes L^{^{\!*}} \big)=h^2\big( S^2C^{^*}\otimes L^{^{\!*}} \big) = 0$. 

\vspace{8pt}
3'. Alternatively we can start with the third exterior power sequence:
\vspace{-4pt}
\beq\label{eq:sequence4}
0\ \longrightarrow\ \wedge^3V \longrightarrow\ \wedge^3B\ \longrightarrow\ \wedge^2B\otimes C\ \longrightarrow\ B\otimes S^2C\ \longrightarrow\ S^3C\ \longrightarrow\ 0
\vspace{-4pt}
\eeq
then split this sequence into three short exact sequences, whose long exact sequences in cohomology lead to the following indentifications:
\vspace{-4pt}
\begin{align}
 \text{Hom}_X\big(L,\wedge^3V\big)& \cong \text{Ker}\left( H^0\big(\wedge^3B\otimes L^{^{\!*}}\big)\ \longrightarrow\ H^0\big(Q_1\big)\right)\\
 H^0(Q_1)&\cong \text{Ker}\left( H^0\big(\wedge^2B\otimes C\otimes L^{^{\!*}}\big)\ \longrightarrow\ H^0\big(Q_2\big)\right)\\
  H^0(Q_2)&\cong \text{Ker}\left( H^0\big(B\otimes S^2C\otimes L^{^{\!*}}\big)\ \longrightarrow\ H^0\big(S^3C\otimes L^{^{\!*}}\big)\right)
  \vspace{-4pt}
\end{align}

\vspace{-8pt}
Thus, if $h^0\big(\!\wedge^3B\otimes L^{^{\!*}}\big)=0$, then $L$ does not inject into $\wedge^3V$. On the other hand, if 
\vspace{-4pt}
\beq\label{eq:cohcondiiton3}
h^0\big(\wedge^3B\otimes L^{^{\!*}}\big)> h^0\big(\wedge^2B\otimes C\otimes L^{^{\!*}}\big)
\vspace{-4pt}
\eeq 
then $\text{Hom}_X\big(L,\wedge^3V\big)$ is non-trivial. 

\vspace{8pt}
4. The case $k=4$, $L\hookrightarrow \wedge^4 V\cong V^{^*}$. In this case, start with the dual monad sequence twisted up with $L^{^{\!*}}$: 
\vspace{-4pt}
\begin{equation} \label{eq:dualmonad}
0\ \longrightarrow\ C^{^*}\otimes L^{^{\!*}} \longrightarrow\ B^{^*}\otimes L^{^{\!*}}\ \longrightarrow\ V^{^*}\otimes L^{^{\!*}}\ \longrightarrow\ 0  
\vspace{-4pt}
\end{equation}
and obtain the long exact sequence in cohomology:
\vspace{-4pt}
\beq
0\ \longrightarrow\ H^0\big( C^{^*}\otimes L^{^{\!*}} \big) \longrightarrow\ H^0\big(B^{^*} \otimes L^{^{\!*}}\big)\ \longrightarrow\ \text{Hom}_X\big(L,V^{^*})\ \longrightarrow\ H^1\big( C^{^*}\otimes L^{^{\!*}} \big) \longrightarrow\ \ldots
\vspace{-4pt}
\eeq
This implies the following identification:
\vspace{-8pt}
\beq
\text{Hom}_X\big(L,V^{^*}) \cong \text{Coker}\left(H^0\big( C^{^*}\otimes L^{^{\!*}} \big) \longrightarrow H^0\big(B^{^*} \otimes L^{^{\!*}}\big)  \right) \oplus \text{Ker}\left(H^1\big( C^{^*}\otimes L^{^{\!*}} \big) \longrightarrow H^1\big(B^{^*} \otimes L^{^{\!*}}\big)  \right)
\vspace{4pt}
\eeq

If $h^0\big(B^{^*} \otimes L^{^{\!*}}\big)=h^1\big( C^{^*}\otimes L^{^{\!*}} \big) = 0$, then $L$ does not inject into $V^{^*}$. However, $\text{Hom}_X\big(L,V^{^*})$ is non-trivial if 
\vspace{-12pt}
\begin{align}\label{eq:cohcondiiton4}
h^0\big( C^{^*}\otimes L^{^{\!*}} \big)&<h^0\big( B^{^*}\otimes L^{^{\!*}} \big) \text{  or }\\ 
\label{eq:cohcondiiton5}
h^1\big( C^{^*}\otimes L^{^{\!*}} \big) &>h^1\big( B^{^*}\otimes L^{^{\!*}} \big)
\vspace{-4pt}
\end{align}

\newpage

\bibliography{bibfile}{}
\bibliographystyle{utcaps}
\end{document}